\documentclass[a4paper,11pt]{article}
\pdfoutput=1  
\usepackage{jcappub}
\usepackage[T1]{fontenc}
\usepackage{float}
\usepackage{hyperref}
\usepackage{multirow}
\usepackage[table]{xcolor}
\usepackage{booktabs}
\usepackage{adjustbox}
\usepackage[normalem]{ulem}
\usepackage{orcidlink}

\title{\boldmath Testing Statistical Isotropy in the FRB Sky Distribution: A Selection-Function-Aware Framework}

\author[a,1]{Bruno W.N. Ribeiro\,\orcidlink{0000-0002-6657-263X}\note{Corresponding author.}}
\author[b]{Thais Lemos\,\orcidlink{https://orcid.org/0009-0002-3365-8418}}
\author[b]{Carlos A.P. Bengaly\,\orcidlink{https://orcid.org/0000-0001-5731-3348}}
\author[c]{Rodrigo Gonçalves\,\orcidlink{https://orcid.org/0000-0002-9189-1261}}
\author[a,d]{K.E.L. de Farias\,\orcidlink{0000-0002-9418-9566}}
\author[a]{Lázaro L. Sales\,\orcidlink{0000-0002-5352-6642}}
\author[a]{Amilcar R. Queiroz\,\orcidlink{0000-0002-4785-5589}}
\author[b]{Jailson Alcaniz\,\orcidlink{https://orcid.org/0000-0003-2441-1413}}

\affiliation[a]{Unidade Acadêmica de Física, Universidade Federal de Campina Grande, Campina Grande, PB, 58429-900, Brazil}
\affiliation[b]{Observatório Nacional, Rio de Janeiro, RJ, 20921-400, Brazil}
\affiliation[c]{Departamento de Física, Universidade Federal Rural do Rio de Janeiro, Seropédica, RJ, 23897-000, Brazil}
\affiliation[d]{Centre of Excellence ENSEMBLE3 Sp. z o. o., Wolczyńska Str. 133, 01-919, Warsaw, Poland}

\emailAdd{bruno\_wesley@uaf.ufcg.edu.br}
\emailAdd{thaislemos@on.br}
\emailAdd{carlosbengaly@on.br}
\emailAdd{rsg\_goncalves@ufrrj.br}
\emailAdd{klecio.lima@uaf.ufcg.edu.br}
\emailAdd{lazaro0254@gmail.com}
\emailAdd{amilcarq@df.ufcg.edu.br}
\emailAdd{alcaniz@on.br}

\abstract{
We perform a test of statistical isotropy in the Universe using the sky distribution of fast radio bursts (FRBs), based on a compilation of $4066$ events detected by multiple surveys. Our method is based on the two-point angular correlation function $w(\theta)$ as in the Landy--Szalay estimator, together with a tomographic absolute-anisotropy statistic, and estimates their observational uncertainties from complementary jackknife and bootstrap resampling. Both estimators are confronted with hierarchical ensembles of isotropic mock catalogs that propagate the uncertainties of empirically reconstructed survey selection functions, as well as the Poisson fluctuations of the isotropic realizations. The significances are obtained from a covariance-aware, SVD-regularized $\chi^2$ statistic calibrated empirically against the mock ensemble, and we evaluate four nested scenarios that progressively incorporate a Galactic-plane mask and the survey selection functions. As for our results, we find that the raw FRB sky is strongly inconsistent with isotropy; Galactic masking alone reduces the tension by only a factor of $\sim 3$, whereas the selection functions reduce it by nearly four orders of magnitude, showing that the apparent anisotropy is driven by the highly non-uniform sky coverage of the contributing surveys, overwhelmingly dominated by CHIME. Only when both effects are combined we obtain that the observed distribution is fully consistent with statistical isotropy. This result is independently corroborated by the absolute-anisotropy estimator, and is stable under variations of the analysis parameters. Therefore, we find that the FRB sky distribution is consistent with statistical isotropy, helping confirm one of the main predictions of the standard model scenario.
}

\begin{document}
\maketitle
\flushbottom


\section{Introduction}

The current standard cosmological model, the $\Lambda$ Cold Dark Matter ($\Lambda$CDM) model, provides an excellent description of observations of the cosmic microwave background (CMB) and the large-scale structure of matter in the Universe \cite{Planck2018}. One of its cornerstones is the so-called Cosmological Principle (CP), which is based on the fundamental assumptions that the Universe is statistically homogeneous and isotropic on sufficiently large scales (see \cite{Goodman1995,Clarkson_2010} for a discussion). This hypothesis allows cosmic distances and ages to be directly derived from the Friedmann-Lemaître-Robertson-Walker (FLRW) metric. Any violation of the CP would require a reformulation of the concordance cosmological model ($\Lambda$CDM). For this reason, testing the assumption of cosmic isotropy with independent tracers of the large-scale Universe is essential to assess the validity of the CP.

Several cosmological observations have been used to test the validity of the CP. For instance, CMB temperature anisotropies \cite{Planck2018_isotropy}, Type Ia supernova (SNe Ia) distances \cite{Andrade_2018a,Andrade_2018b,Bengaly_2024,Mokeddem_2025}, gamma-ray bursts (GRBs) \cite{Andrade_2019,Lopes_2024}, galaxy number counts \cite{Bengaly_2018b,Franco_2023}, and galaxy clusters \cite{Bengaly_2016} have provided evidence that the CP holds in terms of statistical isotropy. However, some observations have revealed statistically significant signals that may challenge the CP, such as the large-angle anomalies in the CMB \cite{Planck2018_isotropy,Kester_2023} and the dipole anisotropy observed in AGNs/quasar source counts observed in radio and infrared \cite{Bengaly_2018a, Secrest_2025, Mittal_2026} (see~\cite{Periv_2022, Aluri_2022} for reviews on this subject). In the context of cosmic homogeneity, there is evidence for the existence of a cosmic homogeneity scale based on galaxy and quasar number count distributions across the Universe~\cite{Ntelis_2017,Shao:2023sxk, Shao:2024qrd, Shao:2025xgi,Lopes2026}, as well as by means of null, cosmological model-independent tests~\cite{LopesDias_2025, Dinda_2025}.

More recently, Fast Radio Bursts (FRBs) \cite{Shen_2026} have also been used as sources for CP tests. These are highly energetic transient events characterized by millisecond durations and radio-frequency emission (typically at GHz frequencies) (for reviews, see \cite{Thornton_2013,Petroff_2014,Petroff_2016}). Their large dispersion measures (DMs) strongly suggest an extragalactic, and often cosmological, origin. The first FRB was detected in 2007 by the Parkes telescope \cite{Lorimer_2007}, and since then, numerous events have been observed by different surveys, particularly by the Canadian Hydrogen Intensity Mapping Experiment (CHIME), which recently released its second catalog containing approximately $4500$ events \cite{CHIMEFRB_2026}. Although the number of detected events has increased significantly in recent years owing to radio telescopes dedicated to FRB searches, only about $100$ FRBs have been well localized, with the corresponding redshift. When combined with their DM, these measurements make FRBs powerful astrophysical and cosmological probes (see \cite{Walters2018,Wei2018,Lin2021,Wu2022,Lemos2025,Sales_2026,Sales_2026b,Ribeiro2026} for cosmological applications of FRBs).

Since the number of detected events is increasing with the new surveys, the FRBs constitute a promising observational probe for testing large-scale isotropy. In the recent work \cite{Shen_2026}, the authors test cosmic isotropy using the first CHIME/FRB catalog, showing results that are consistent with isotropy. However, in this case, the statistics are limited by the observational sample, which covers only the northern hemisphere and contains $\sim$ 500 FRBs, as the authors used only the CHIME first FRB catalog. 

In the present work, we expand the analysis of \cite{Shen_2026} by testing the statistical isotropy of the Universe using the sky distribution of the current FRB sample. Our analysis is based on a compilation of $4066$ FRBs detected by multiple surveys across the entire sky, and we consider three additional scenarios: applying a Galactic plane mask, which restricts the sample to the $2518$ FRBs with $|b|>20^\circ$; applying a selection-function criterion specific to each survey; and combining both approaches. We then test the isotropy of each sample using two complementary estimators: the two-point angular correlation function (2pACF), calculated with the Landy--Szalay estimator, and the tomographic absolute-anisotropy statistic. Their observational uncertainties are estimated through jackknife and bootstrap resampling, while both estimators are tested against ensembles of isotropic mock catalogs that account for uncertainties in the empirically reconstructed survey selection functions and Poisson fluctuations. The statistical significance is assessed using a covariance-aware statistical framework calibrated against these mock realizations.

This paper is organized as follows. We present the sample data used in our analysis in Sec.~\ref{sec:data}. In Sec.~\ref{sec:methodology}, we describe the methodology and the data analysis performed. We present and discuss our main results in Sec.~\ref{sec:results}. Finally, in Sec.~\ref{sec:conclusion}, we summarize the conclusions.


\section{The Observational Data Set}
\label{sec:data}

For the present analysis of the statistical isotropy of the FRB sky distribution, we compile the FRBs detected by multiple surveys. The main source is the Transient Name Server (TNS) database\footnote{\url{https://www.wis-tns.org}}, which includes the second CHIME/FRB catalog with $\sim 4500$ events \cite{CHIMEFRB_2026}, Commensal Real-time ASKAP Fast Transient (CRAFT) \cite{Shannon_2025}, and Deep Synoptic Array-110 (DSA-110) \cite{Connor_2024}. We complement this sample with observations from the Five-hundred-meter Aperture Spherical radio Telescope (FAST) \cite{FAST} data available at \footnote{\url{https://blinkverse.zero2x.org/}}; and the Parkes telescope datasets \cite{Yang_2025}. 

Since our goal is to test statistical isotropy signal from FRBs, we remove the following events: (i) FRBs detected by the DSA-110 survey, whose highly non-uniform sky distribution, with all events concentrated in the northernmost region of the sky, could bias the signal; (ii) CHIME events whose reported flux density lies below the telescope's minimum detection sensitivity \footnote{We apply this sensitivity selection criterion only to the CHIME survey, because this information is not provided for the other surveys, such as the observed flux density of each event and the telescope's minimum detection sensitivity.}; (iii) and the FRBs for which the reporting survey is not available and therefore cannot be assigned a selection function (described in the next section). The resulting working sample contains $4066$ events and is used in the scenarios without Galactic masking. When the Galactic Plane mask is applied, the fiducial cut $|b|>20^\circ$ further reduces the sample to $2518$ FRBs. For each burst, we use the following main properties: (i) RA, the Right Ascension of the burst, given in J2000 decimal degrees; (ii) DEC, the Declination of the burst, also given in J2000 decimal degrees. It is worth noting that the positional uncertainties are unavailable for several FRBs in the catalogues. Therefore, we employ the jackknife and bootstrap resampling techniques to estimate the variance of our analysis. Both methods are described in the next section. 

\begin{figure*}
\begin{center}
\includegraphics[width=0.95\textwidth]{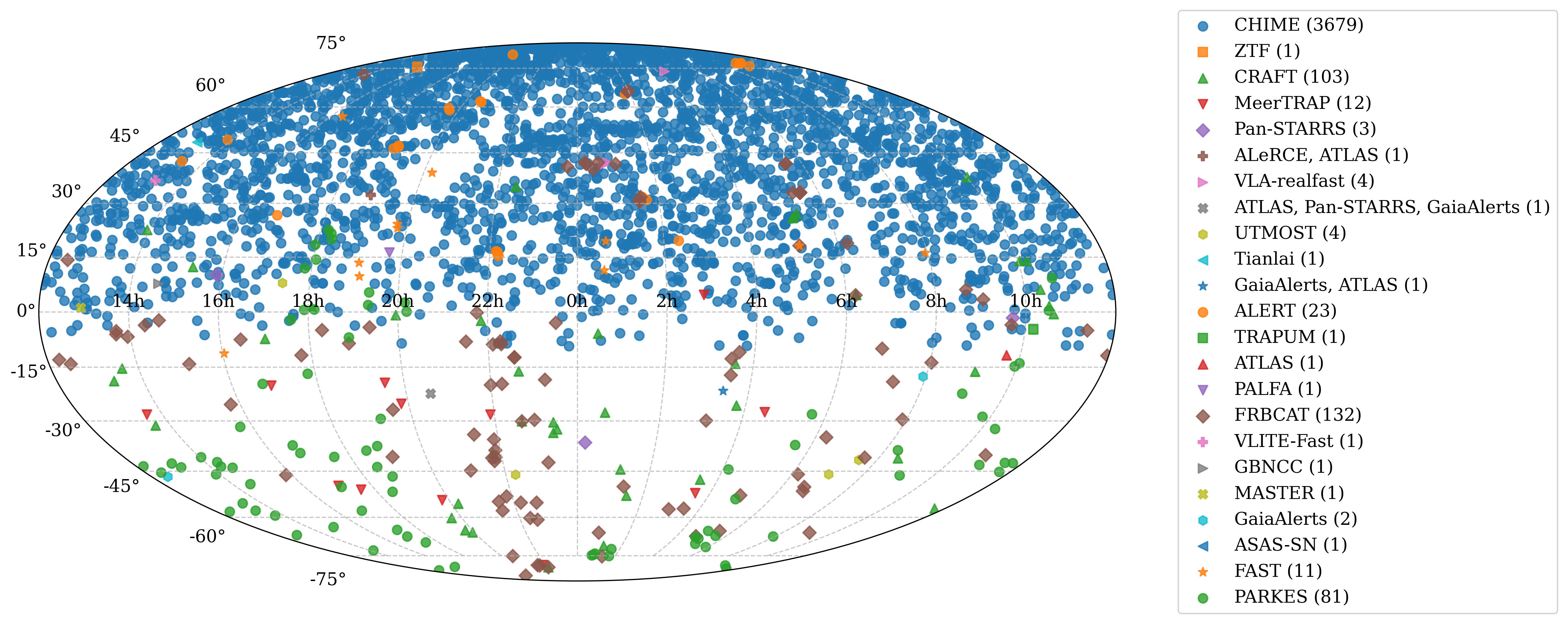}
\caption{Mollweide projection of the sky positions of FRBs from different surveys. The plot was generated using the \texttt{matplotlib.pyplot} package.}
\label{fig:skymap_frbs}
\end{center} 
\end{figure*}

The sky distribution of the selected FRBs is displayed in Fig.~\ref{fig:skymap_frbs}. Note that there are significantly more observed events in the northern hemisphere ($\sim 3800$) than in the southern hemisphere ($\sim 250$), mainly due to the CHIME survey, whose field of view covers the entire northern sky and extends slightly into the southern hemisphere ($0^\circ \lesssim \mathrm{DEC} \lesssim 90^\circ$ in our sample), and was specifically designed for transient radio searches, including FRBs. The CHIME Collaboration has also reported a decrease in the number of detections near the Galactic plane, which can be seen in Fig.~\ref{fig:skymap_frbs}. As a result, these observational selection effects may introduce a bias in tests of statistical isotropy, which we attempt to address with our methodology, as explained further.

\section{Methodology}
\label{sec:methodology}
 
\begin{figure*}
\centering
\includegraphics[width=0.95\textwidth]{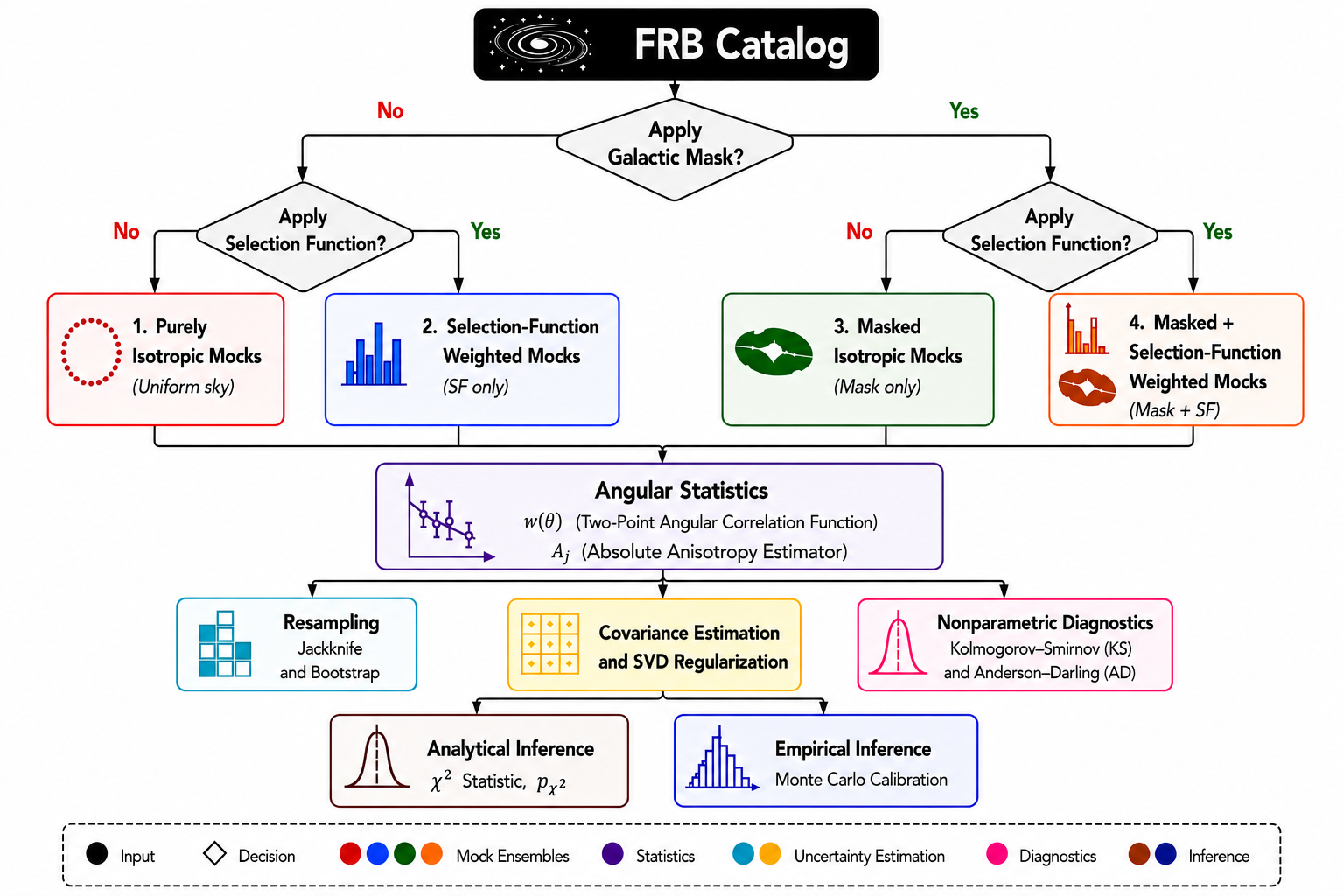}
\caption{Schematic overview of the analysis pipeline. The FRB catalog is subjected to Galactic masking and empirical survey selection, yielding four isotropic mock configurations. The observed and mock catalogs are analyzed through the Landy--Szalay two-point angular correlation function $w(\theta)$ and the absolute anisotropy estimator $A_j$, followed by resampling, covariance estimation and regularization, nonparametric diagnostics, and analytical and empirical inference. The masked plus selection-function weighted ensemble defines the fiducial observational null hypothesis.}
\label{fig:pipeline}
\end{figure*}

In this section, we describe the methodology designed to test the hypothesis of statistical isotropy of the Universe in light of the observed FRB angular distribution presented in Sec.~\ref{sec:data}. The framework combines Galactic masking, empirical survey selection functions, hierarchical isotropic mock ensembles, covariance modeling, and covariance-aware statistical inference, evaluating whether the observed FRB distribution retains residual anisotropies beyond those expected from this empirically reconstructed observational model. Since publicly available survey information is insufficient to construct a consistent angular exposure model across all surveys, we estimate the survey selection functions (SFs) directly from the observed sky distributions, i.e., via an empirical approach. As a consequence, the inferred selection functions may partially absorb genuine large-scale anisotropy signals, as expected in typical raw observational data. A schematic overview of the complete pipeline is presented in Fig.~\ref{fig:pipeline}.

The pipeline is implemented in \texttt{Python}\footnote{The full analysis code is publicly available at \url{https://github.com/bwesley92/FRB-isotropy-tests}.}, relying on \texttt{astropy} \cite{Astropy_2013} for equatorial-to-Galactic coordinate transformations, \texttt{healpy} \cite{Gorski_2005} for spherical pixelization, \texttt{TreeCorr} \cite{Jarvis_2004} for angular pair counting, and \texttt{scikit-learn} \cite{Pedregosa_2011} for the shrinkage covariance estimator. All fiducial parameters are introduced at their first appearance and consolidated in Appendix~\ref{app:fiducial} for quick reference.

\subsection{Sky Pixelization and Galactic Masking}
\label{subsec:mask}
 
The celestial sphere is discretized using the HEALPix formalism \cite{Gorski_2005}, which provides an equal-area spherical pixelization optimized for large-scale sky analyses. All angular maps employed in this work, i.e., survey selection functions, jackknife regions, and overlap diagnostics, are constructed in HEALPix space, and each FRB is assigned to its pixel through the standard angular transformation $p = {\rm ang2pix}(N_{\rm side},\theta,\phi)$, where $(\theta,\phi)$ are the spherical coordinates of the source position. We adopt a fiducial resolution $N_{\rm side}^{\rm SF}=32$ for the selection-function reconstruction, and a coarser grid $N_{\rm side}^{\rm JK}=4$ for the jackknife partitioning, chosen to guarantee sufficiently populated regions.
 
Before any other step, the observed equatorial coordinates (RA, DEC) of each FRB are converted to the Galactic frame $(l, b)$, and a latitude mask is applied to suppress regions heavily contaminated by Galactic foregrounds --- primarily driven by dispersion, scattering, sky temperature, source confusion, and survey incompleteness --- which results in strong observational incompleteness. Throughout this work, we adopt the fiducial cut
\begin{equation}
    |b| > b_{\rm cut},
    \qquad
    b_{\rm cut} = 20^\circ.
\end{equation}
The same masking operation is consistently propagated into the selection-function reconstruction, the random and mock catalogs, the jackknife resampling, and the covariance estimators, so that the fiducial null hypothesis is defined only within the observable sky region after masking.

\subsection{Survey Selection Functions}
\label{subsec:sf}

The catalog is partitioned according to the reporting survey of each FRB, and an important because the observed sky distribution is not determined solely by the underlying FRB population, but is also shaped by the distinct observational characteristics of each survey, including its field of view, sky coverage, sensitivity, and observing strategy. Combining FRBs from different surveys without accounting for their individual selection effects could introduce artificial anisotropies into the observed distribution. For this reason, to take into account these effects, we construct an angular selection function of each survey from the spatial distribution of its detected FRBs. This is achieved by constructing a sky count map for each survey, in which the observed FRBs are used to characterize their corresponding angular distribution. For each survey, a sky count map is built by assigning every FRB to its HEALPix pixel at resolution $N_{\rm side}^{\rm SF}$. The raw selection function is
\begin{equation}
    {\rm SF}(p) =
    \frac{N(p)}{\sum_p N(p)},
\end{equation}
where $N(p)$ is the number of FRBs in pixel $p$. By construction, ${\rm SF}(p)$ is normalized to unity and represents the probability that an FRB associated with a given survey is detected in the direction of pixel $p$.
 
Because current FRB samples remain sparse, the raw pixelized maps are strongly affected by shot noise at small angular scales. To recover the large-scale observational response, each selection function is smoothed with a Gaussian kernel,
\begin{equation}
{\rm SF}{\rm smooth}(\hat n)
=
G_{\sigma_{\rm smooth}} \ast {\rm SF}(\hat n),
\end{equation}
where $\ast$ denotes spherical convolution, $G_{\sigma_{\rm smooth}}$ is the corresponding Gaussian kernel with smoothing scale $\sigma_{\rm smooth}$, and $\hat{n}$ is the position of the pixel $p$. We adopt the fiducial value $\sigma_{\rm smooth}=3^\circ$, a few times the native HEALPix pixel scale at the fiducial resolution $N_{\rm side}^{\rm SF} = 32$, implying $\theta_{\rm pix} \approx 1.8^\circ$, large enough to average over pixel-to-pixel Poisson shot noise, yet nearly two orders of magnitude below the declination range over which genuine large-scale survey-footprint gradients occur. To remove the angular leakage across mask boundaries induced by the convolution, the Galactic mask is reapplied to the smoothed map, followed by a final renormalization.

The total observational selection model is then a weighted mixture of the independent survey selection functions, with statistical weight
\begin{equation}
    w_i =
    \frac{N_i}{\sum_j N_j},
\end{equation}
where $N_i$ is the number of FRBs associated with survey $i$. FRBs reported by multiple surveys contribute independently to each corresponding $N_i$.

\subsection{Hierarchical Mock Ensembles}
\label{subsec:mocks}

Four distinct classes of isotropic realizations are considered throughout this work, defined by whether the Galactic mask and/or the empirical survey selection functions are applied:
\begin{enumerate}
    \item \textit{Pure isotropy}: ideal isotropic skies generated from a perfectly uniform angular distribution;
    \item \textit{Masked isotropy}: isotropic skies masked through the removal of the Galactic plane region;
    \item \textit{SF-weighted isotropy}: isotropic skies filtered through the empirical survey selection functions;
    \item \textit{Fiducial}: isotropic skies simultaneously masked and filtered through the empirical survey selection functions.
\end{enumerate}
Each configuration is applied identically to the observed catalog and to its corresponding hierarchical mock ensemble: each mock isotropic catalog is drawn with the same number of objects as its corresponding observed catalog ($4066$ for the two unmasked scenarios and $2518$ for the two masked ones) so that observed and simulated data are always compared on equal observational footing. Together, these four configurations isolate and quantify the individual and combined impacts of Galactic masking and instrumental selection on the inferred angular isotropy. The fourth class defines the primary observational null hypothesis tested in this work, namely whether the observed FRB distribution is statistically compatible with an intrinsically isotropic population after propagation through the observational response of the surveys.
 
The Selection Function uncertainties are propagated through the ensemble construction. Assuming Poisson statistics, the uncertainty of each HEALPix pixel is approximated by $\sigma_N(p)=\sqrt{N(p)}$, with diagonal covariance contribution
\begin{equation}
    {\rm Cov}_{\rm SF}(p) =
    \left(\frac{\sigma_N(p)}{N_{\rm total}}\right)^{\!2},
\end{equation}
evaluated directly from the unsmoothed pixel counts, since the dominant source of uncertainty is shot noise at the native pixel scale. Perturbed realizations of each survey selection function are then drawn as ${\rm SF}_{\rm pert}(p) = {\rm SF}(p) + \delta {\rm SF} (p)$, where $\delta {\rm SF}(p)$ is a Gaussian variate with zero mean and standard deviation $\sqrt{{\rm Cov}_{\rm SF}(p)}$. Non-positive values are clipped at zero, and the perturbed map is renormalized to unit sum.
 
Each perturbed SF realization defines one ensemble-level observational response, and multiple statistically independent mock catalogs are generated conditioned on that fixed realization. The mock hierarchy therefore simultaneously propagates instrumental uncertainties in the empirical SFs and intrinsic Poisson fluctuations of the isotropic realizations. In the fiducial configuration, we generate $N_{\rm ens}=20$ SF realizations and $N_{\rm mocks}^{\rm ens}=50$ isotropic mock catalogs per realization, for a total of $N_{\rm mocks}=N_{\rm ens}\times N_{\rm mocks}^{\rm ens}=1000$ mock catalogs available for covariance estimation and empirical significance calibration.
 
Individual mock events are generated by first drawing a survey label according to its statistical weight $w_i$, and then sampling a HEALPix pixel from the corresponding perturbed selection function, $P(\hat n)\propto {\rm SF}_{\rm pert}(\hat n)$. To avoid artificial angular quantization at the pixel scale, sampled positions are randomized within the solid angle of each pixel: a uniform subpixel offset is drawn in $[-r_{\rm pix},+r_{\rm pix}]$, where $r_{\rm pix}$ is the maximum pixel radius, and the right-ascension offset is rescaled by $1/\cos\delta$ to preserve angular scales near the poles. This restores a continuous angular distribution while preserving the large-scale selection-function structure. For each isotropic realization, an independent random catalog is generated using the same observational selection model, including Galactic masking, survey weighting, and angular selection functions; its size and role within the two-point correlation estimator are detailed in the following subsection.

\subsection{Angular Statistics and Uncertainty Estimation}
\label{subsec:stats}
 
The angular two-point correlation function (2pACF) is computed according to the Landy--Szalay estimator \cite{Landy_1993},
\begin{equation}
    w(\theta) =
    \frac{DD(\theta) - 2DR(\theta) + RR(\theta)}{RR(\theta)},
\end{equation}
where $DD$, $DR$, and $RR$ denote the normalized data--data, data--random, and random--random pair counts, with the random catalog enlarged to $N_{\rm rand}/N_{\rm data}=20$ to suppress Monte Carlo (MC) shot noise in the pair counts. Pair counting is performed with \texttt{TreeCorr} \cite{Jarvis_2004} in linear angular bins spanning $0^\circ \le \theta \le 180^\circ$, with a fiducial bin width $\Delta\theta=10^\circ$, i.e., $N_{\rm bins}=18$ fine--grained angular bins, by the same token of~\cite{Andrade_2019}. Fixed nominal bin edges are used across all mock realizations to guarantee a consistent covariance structure.

In addition, we compute a coarse-grained absolute anisotropy estimator, following the tomographic form of \cite{Andrade_2019},
\begin{equation}
    A_j \equiv \sum_{\theta_i\,\in\,\text{bin }j} |w(\theta_i)|,
    \qquad
    j = 1,\dots,N_{\rm tomo},
    \label{eq:abs-sum}
\end{equation}
which avoids cancellations between positive and negative fluctuations. In practice, the fine-grained profile $w(\theta_i)$ is first partitioned into $N_{\rm tomo}=9$ coarser tomographic bins of width $\Delta\theta_{\rm tomo}=20^\circ$, spanning $0^\circ$--$180^\circ$, and the absolute values of $w(\theta_i)$ falling within each tomographic bin $j$ are summed to form the tomographic profile. The global anisotropy statistic is then obtained by summing over all tomographic bins, $A \equiv \sum_{j=1}^{N_{\rm tomo}} A_j$, which measures the cumulative angular anisotropy power.

The statistical uncertainties associated with both $w(\theta)$ and $A_j$ are estimated using jackknife \cite{Norberg_2009} and bootstrap \cite{Tarnopolski_2017} resampling of the observed catalog. The jackknife procedure follows a leave-one-region-out strategy based on HEALPix regions with $N_{\rm side}^{\rm JK}=4$, whereas the bootstrap is constructed from resampled catalogs generated with replacement. As with the mock ensemble, the jackknife and bootstrap procedures are controlled by their own fixed, independent random seeds, so that both resampling estimates are deterministically reproducible for a given configuration.

\subsection{Covariance Modeling and Statistical Inference}
\label{subsec:inference}

The isotropic mock ensemble provides both the reference isotropic benchmark, defined as the ensemble mean of the angular statistics, and the covariance matrix adopted throughout the statistical inference. The covariance is estimated directly from the mock ensemble, regularized using the Ledoit--Wolf shrinkage estimator \cite{Ledoit_2004}, corrected using the Hartlap factor \cite{Hartlap_2006}, and further stabilized through a singular value decomposition (SVD) truncation\footnote{For a symmetric covariance matrix, the SVD is equivalent to an eigendecomposition, factorizing it into an orthogonal basis of eigenmodes ranked by their eigenvalues $\lambda_i$; this provides a numerically robust way to identify and discard poorly constrained, noise-dominated directions before the covariance is inverted \cite{Press_2007}.}. In the latter case, only eigenmodes satisfying $\lambda_i>10^{-2}\lambda_{\max}$ are retained, where $\lambda_{\max}$ is the largest covariance eigenvalue. The number of retained eigenmodes is denoted by $N_{\rm kept}$.

The consistency between the observed catalog and the isotropic benchmark is evaluated using both the full-covariance statistic and its SVD-regularized counterpart \cite{Press_2007}:
\begin{equation}
    \chi^2_{\rm red} = \frac{\Delta^{\rm T} \mathbf{C}_{\rm corr}^{-1}\,\Delta}{N_{\rm bins}},
    \qquad
    \chi^2_{\rm SVD,red} = \frac{\alpha_{\rm SVD}}{N_{\rm kept}}\sum_{i\,\in\,{\rm kept}}\frac{(\mathbf{v}_i^{\rm T}\Delta)^2}{\lambda_i},
\end{equation}
where $\mathbf{C}_{\rm corr}^{-1}$ is the regularized inverse covariance, $\Delta \equiv w_{\rm obs}-\langle w_{H_0}\rangle$ is the residual vector, $\mathbf{v}_i$ are the retained eigenvectors of the covariance matrix, and $\alpha_{\rm SVD}$ is the Hartlap correction. For each statistic, analytical $p$-values are obtained from the corresponding $\chi^2$ distribution,
\begin{equation}
    p_{\rm an} = P(\chi^2\ge\chi^2_{\rm obs}),
    \qquad
    p_{\rm an}^{\rm SVD} = P(\chi^2_{\rm SVD}\ge\chi^2_{{\rm SVD,obs}}),
\end{equation}
while empirical Monte Carlo calibrations are computed from the isotropic mock ensemble,
\begin{equation}
    p_{\rm emp} = \frac{N(\chi_k^2\ge\chi_{\rm obs}^2)+1}{N_{\rm mocks}+1},
    \qquad
    p_{\rm emp}^{\rm SVD} = \frac{N(\chi^2_{{\rm SVD},k}\ge\chi^2_{{\rm SVD,obs}})+1}{N_{\rm mocks}+1}.
\end{equation}

For convenience, the empirical significances are also expressed as Gaussian-equivalent significances,
\begin{equation}
    \sigma_{\chi^2} = \Phi^{-1}(1-p_{\rm emp}),
    \qquad
    \sigma_{\chi^2_{\rm SVD}} = \Phi^{-1}(1-p_{\rm emp}^{\rm SVD}),
\end{equation}
where $\Phi^{-1}$ is the inverse standard-normal cumulative distribution. Throughout this work, the SVD-regularized, empirically-calibrated quantities $\chi^2_{\rm SVD,red}$, $p_{\rm emp}^{\rm SVD}$, and $\sigma_{\chi^2_{\rm SVD}}$ are adopted as the primary isotropy diagnostics, while the corresponding full-covariance quantities are reported for comparison.

The same empirical Monte Carlo calibration is applied to the absolute anisotropy estimator $A$ (Eq.~\ref{eq:abs-sum}), evaluated identically on the observed catalog and on every mock realization,
\begin{equation}
    p_{\rm emp}(A) = \frac{N(A_k\ge A_{\rm obs})+1}{N_{\rm mocks}+1}.
\end{equation}
As a complementary, purely Gaussian measure of deviation, assuming an approximately Gaussian spread of $A$ across mocks, rather than relying on its empirical rank, we further report the tension
\begin{equation}
    \text{tension} \equiv
    \frac{|A_{\rm obs}-\overline{A}_{\rm mocks}|}{\mathrm{std}(A_{\rm mocks})}.
\end{equation}
Unlike the covariance-aware $\chi^2$ statistics, the global estimator carries no explicit angular-bin covariance information. Therefore, it is treated throughout this work as an independent cross-check on the SVD-regularized diagnostic rather than as a replacement for it.

\subsection{Nonparametric Diagnostics}
\label{subsec:nonpara-diagn}

As a complementary diagnostic, we compare the observed $w(\theta)$ and $A_j$ profiles with their corresponding isotropic benchmarks using the two-sample Kolmogorov--Smirnov (KS) and Anderson--Darling (AD) tests \cite{Ivezic_2014,Scholz_1987}. Because the angular and tomographic bins are strongly correlated, we caution that the analytical $p$-values associated with these tests should be regarded only as heuristic. We therefore complement them with empirical Monte Carlo $p$-values estimated directly from the isotropic mock ensemble by comparing the observed statistics with their distributions across the mock realizations; the KS and AD statistics are computed using the \texttt{ks\_2samp} and \texttt{anderson\_ksamp} routines of \textsc{SciPy}.


\section{Results and Discussion}
\label{sec:results}
 
In this section, we present and discuss the results of our main and robustness analyses. The main analysis reports the results of the isotropy diagnostics in the four complementary scenarios of Sec.~\ref{subsec:mocks}, whereas the robustness analysis investigates the stability of these results in relation to our main modeling and methodology choices.

\subsection{Main analysis}
 
\begin{figure}[p]
\centering
\includegraphics[width=0.8\textwidth]{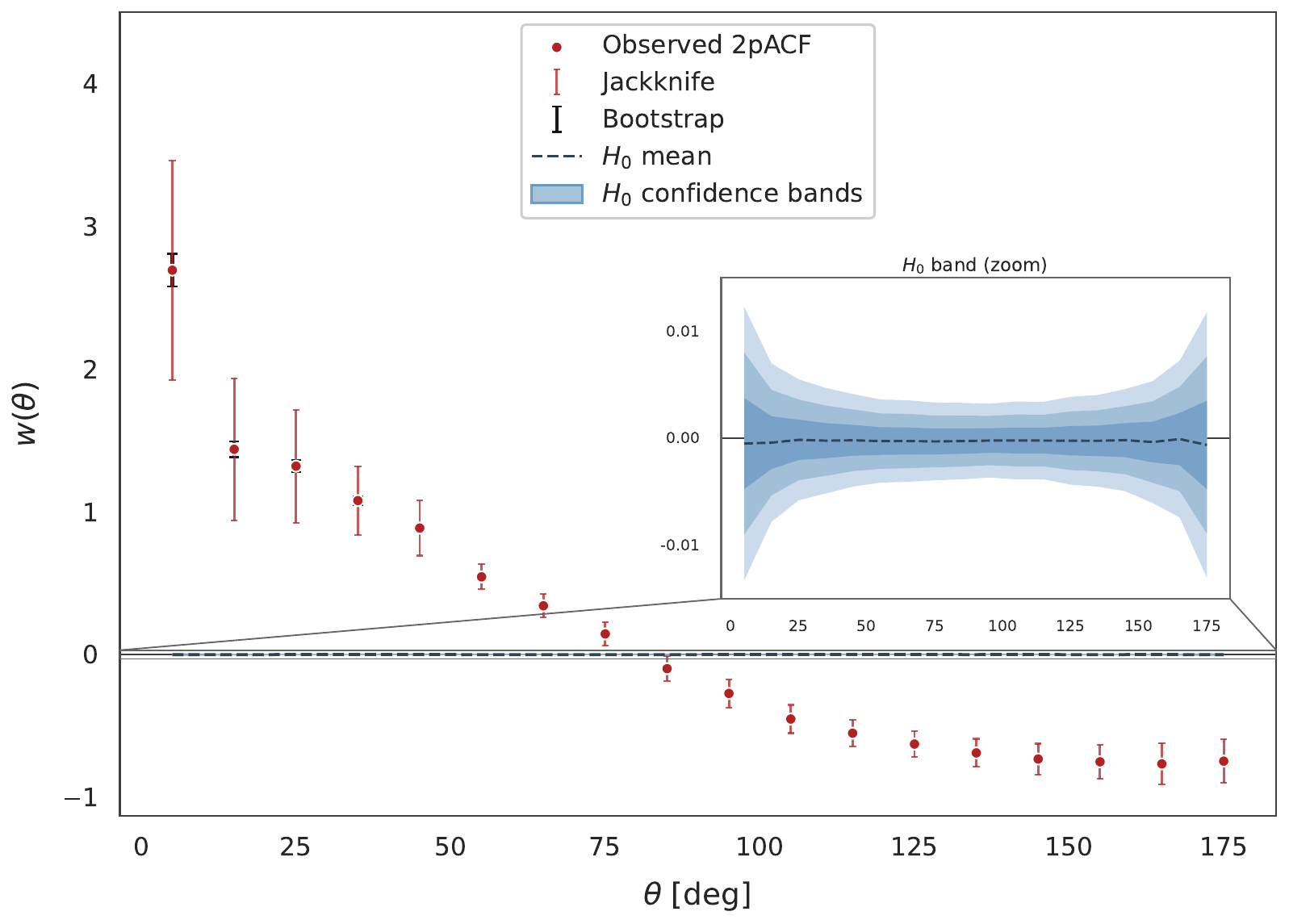}
\caption{2pACF $w(\theta)$ for the \emph{Pure isotropy} scenario. Points show the observed catalog, with jackknife (red) and bootstrap (black) error bars; the dashed line and the shaded bands show, respectively, the mean and the $1\sigma$, $2\sigma$, and $3\sigma$ confidence intervals. Moreover, the $H_0$ mean and confidence bands denote the null hypothesis of statistical isotropy around $w(\theta) = 0$ given Poisson noise and sky selection effects; the inset shows the band magnified around $w(\theta) = 0$, where it is otherwise imperceptible at the scale of the observational signal.}
\label{fig:pure_2pacf}
\vspace{\floatsep}
\includegraphics[width=0.8\textwidth]{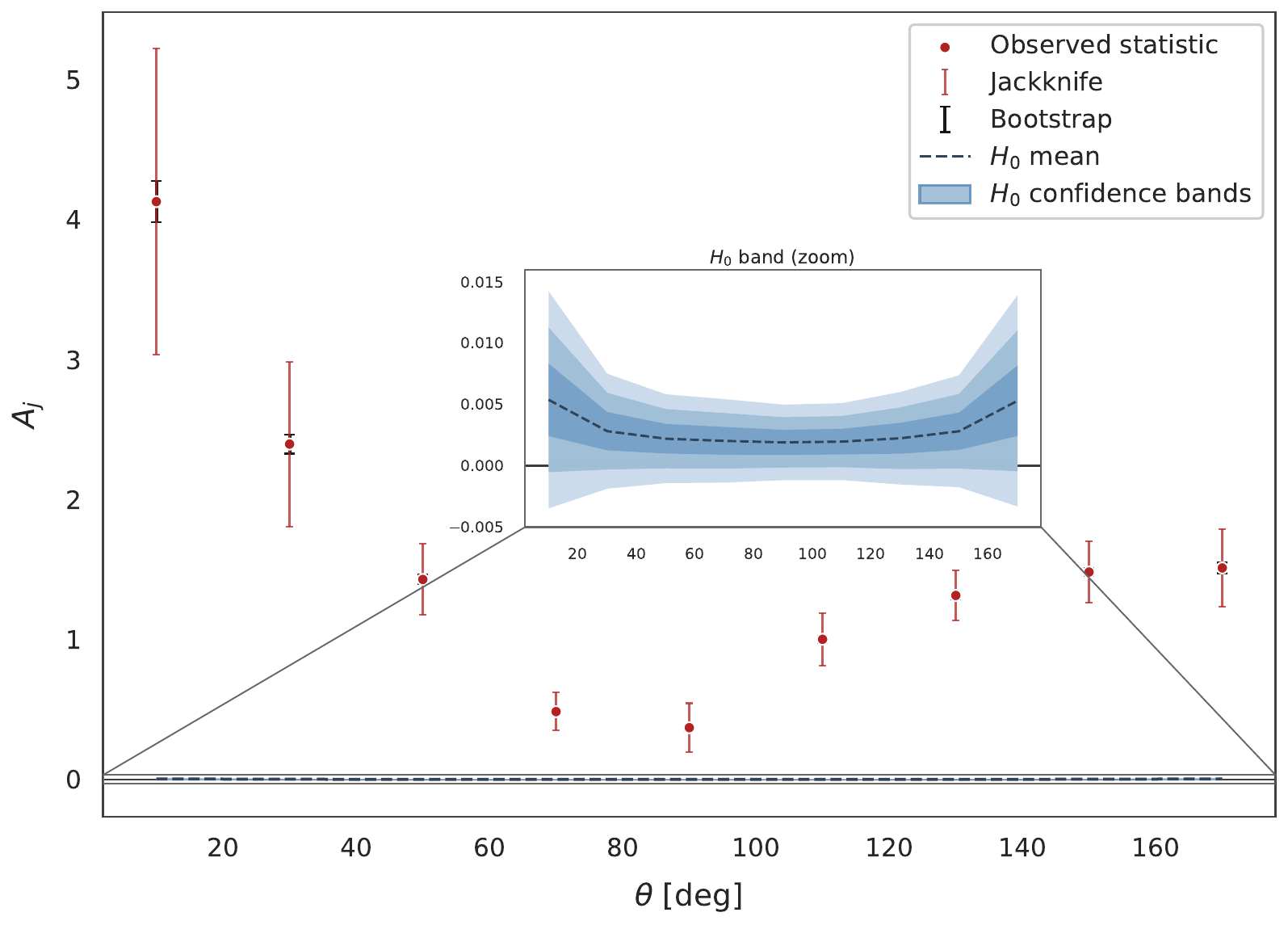}
\caption{Absolute-anisotropy statistic $A_j$ for the \emph{Pure isotropy} scenario. Symbols, error bars, and confidence bands as in Fig.~\ref{fig:pure_2pacf}.}
\label{fig:pure_abs}
\end{figure}

\begin{figure}[p]
\centering
\includegraphics[width=0.8\textwidth]{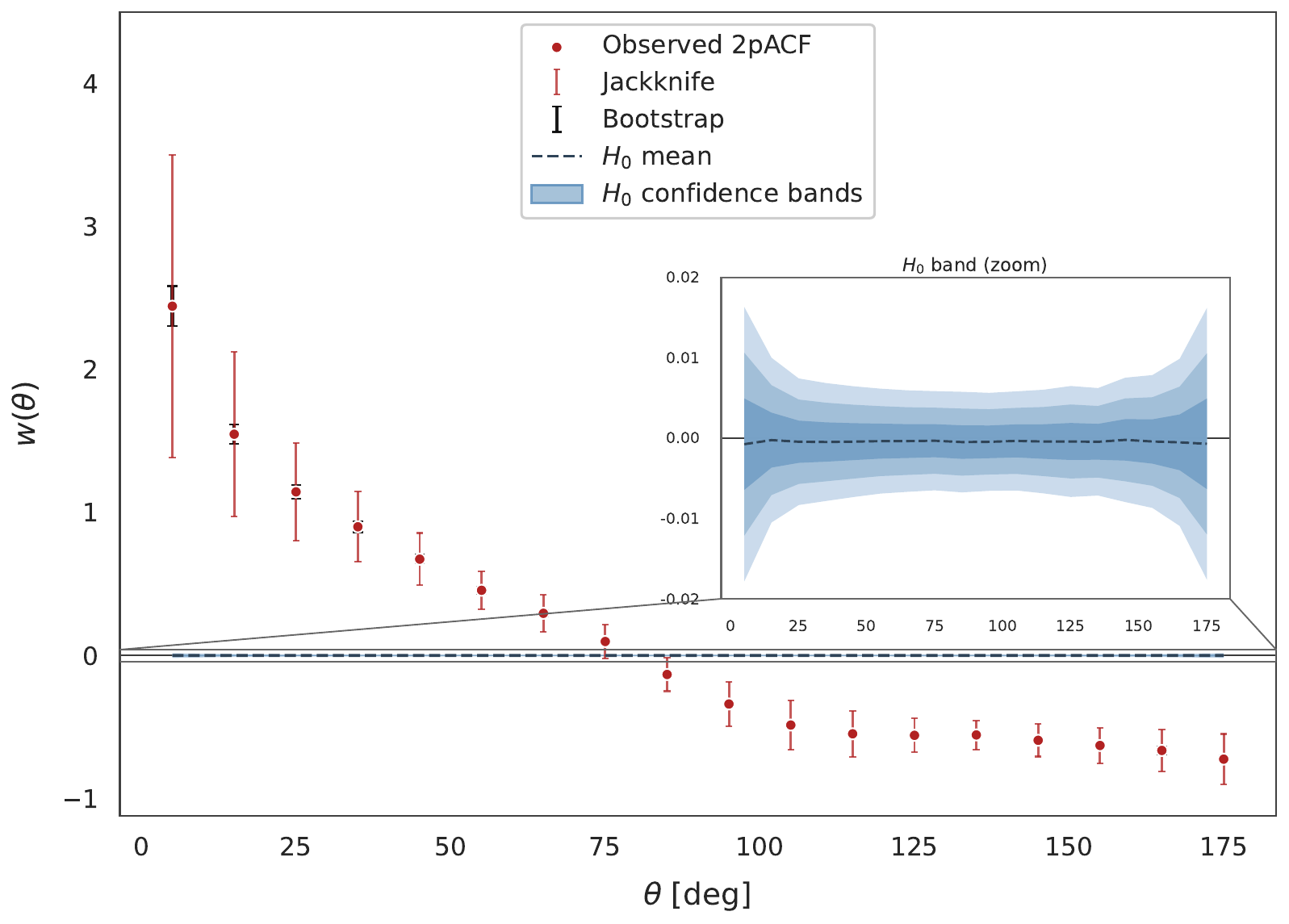}
\caption{Same as Fig.~\ref{fig:pure_2pacf}, but for the \emph{Masked isotropy} scenario.}
\label{fig:masked_2pacf}
\vspace{\floatsep}
\includegraphics[width=0.8\textwidth]{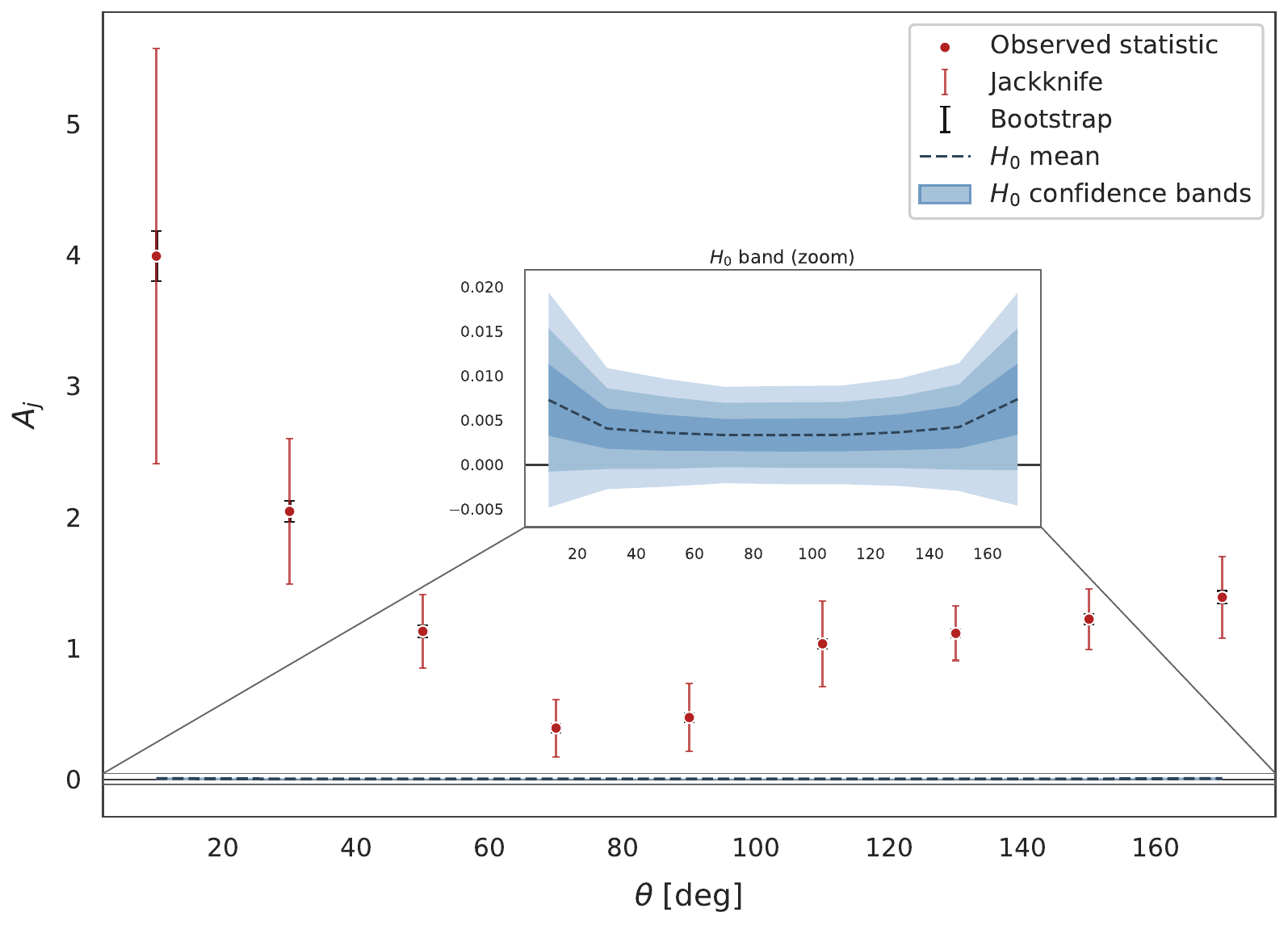}
\caption{Same as Fig.~\ref{fig:pure_abs}, but for the \emph{Masked isotropy} scenario.}
\label{fig:masked_abs}
\end{figure}

\begin{figure}[p]
\centering
\includegraphics[width=0.8\textwidth]{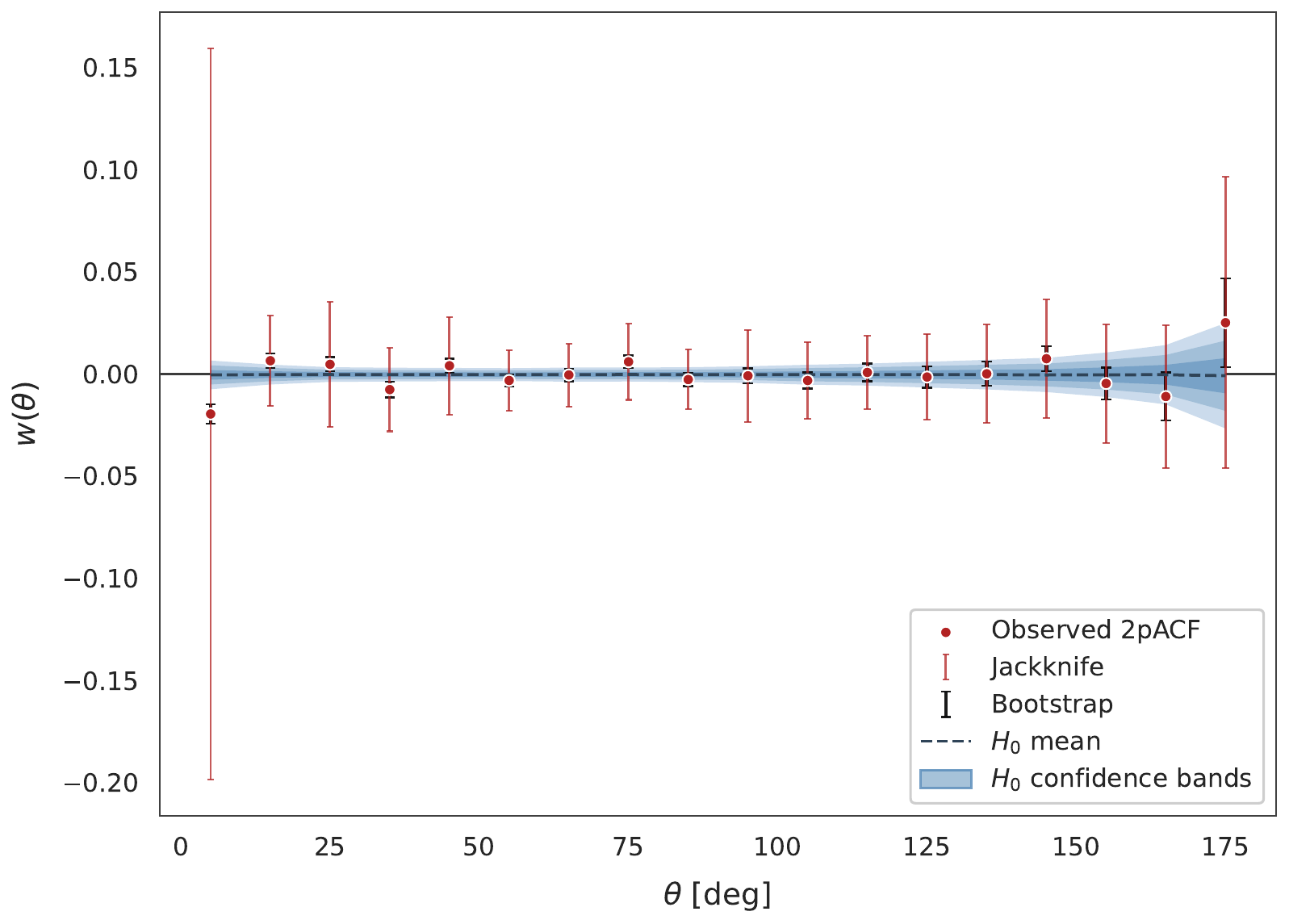}
\caption{Same as Fig.~\ref{fig:pure_2pacf}, but for the \emph{SF-weighted} scenario.}
\label{fig:weighted_2pacf}
\vspace{\floatsep}
\includegraphics[width=0.8\textwidth]{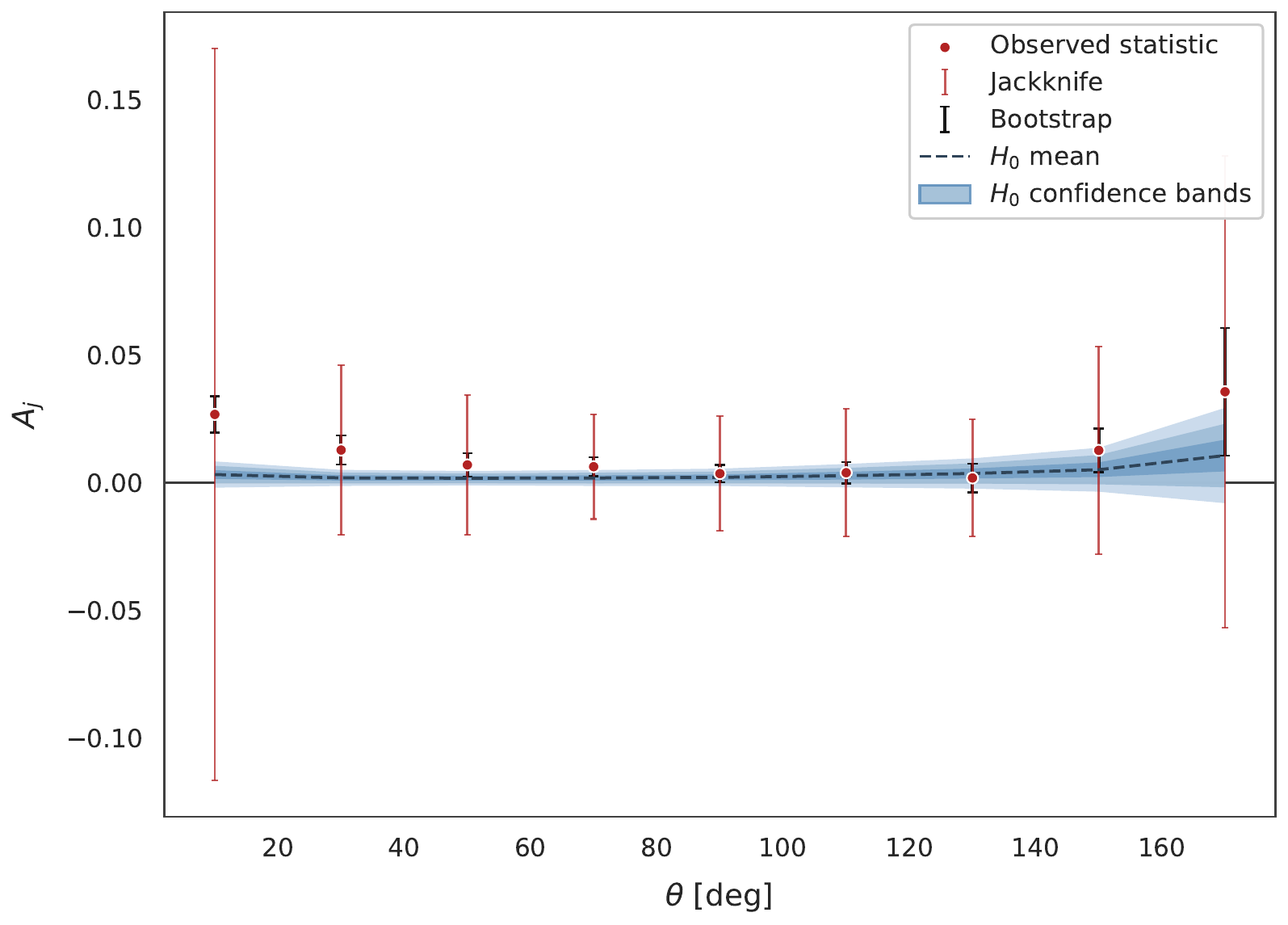}
\caption{Same as Fig.~\ref{fig:pure_abs}, but for the \emph{SF-weighted} scenario.}
\label{fig:weighted_abs}
\end{figure}

\begin{figure}[p]
\centering
\includegraphics[width=0.8\textwidth]{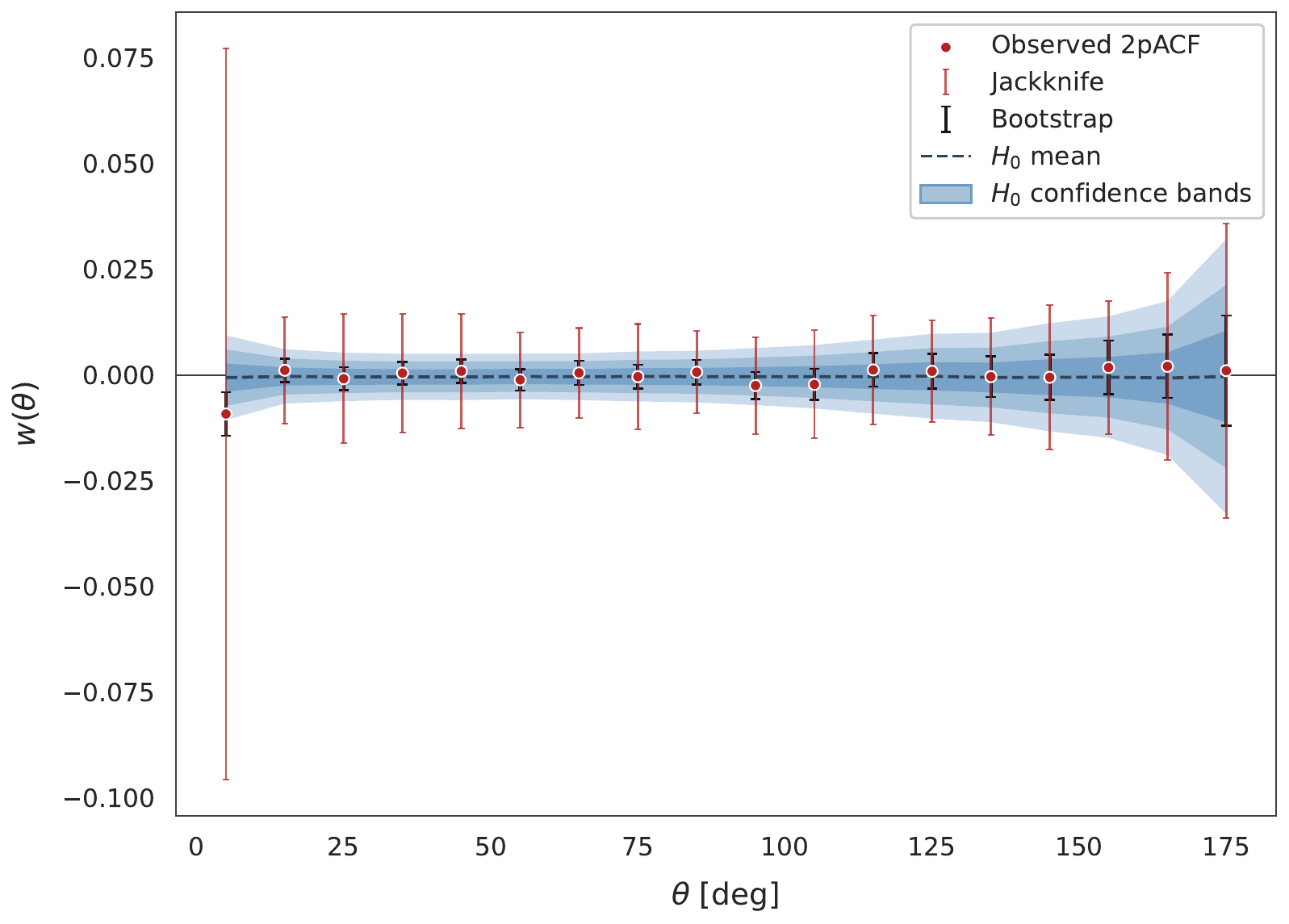}
\caption{Same as Fig.~\ref{fig:pure_2pacf}, but for the \emph{Fiducial} scenario. The observed $w(\theta)$ is statistically consistent with the isotropic mock ensemble at all angular separations.}
\label{fig:fiducial_2pacf}
\vspace{\floatsep}
\includegraphics[width=0.8\textwidth]{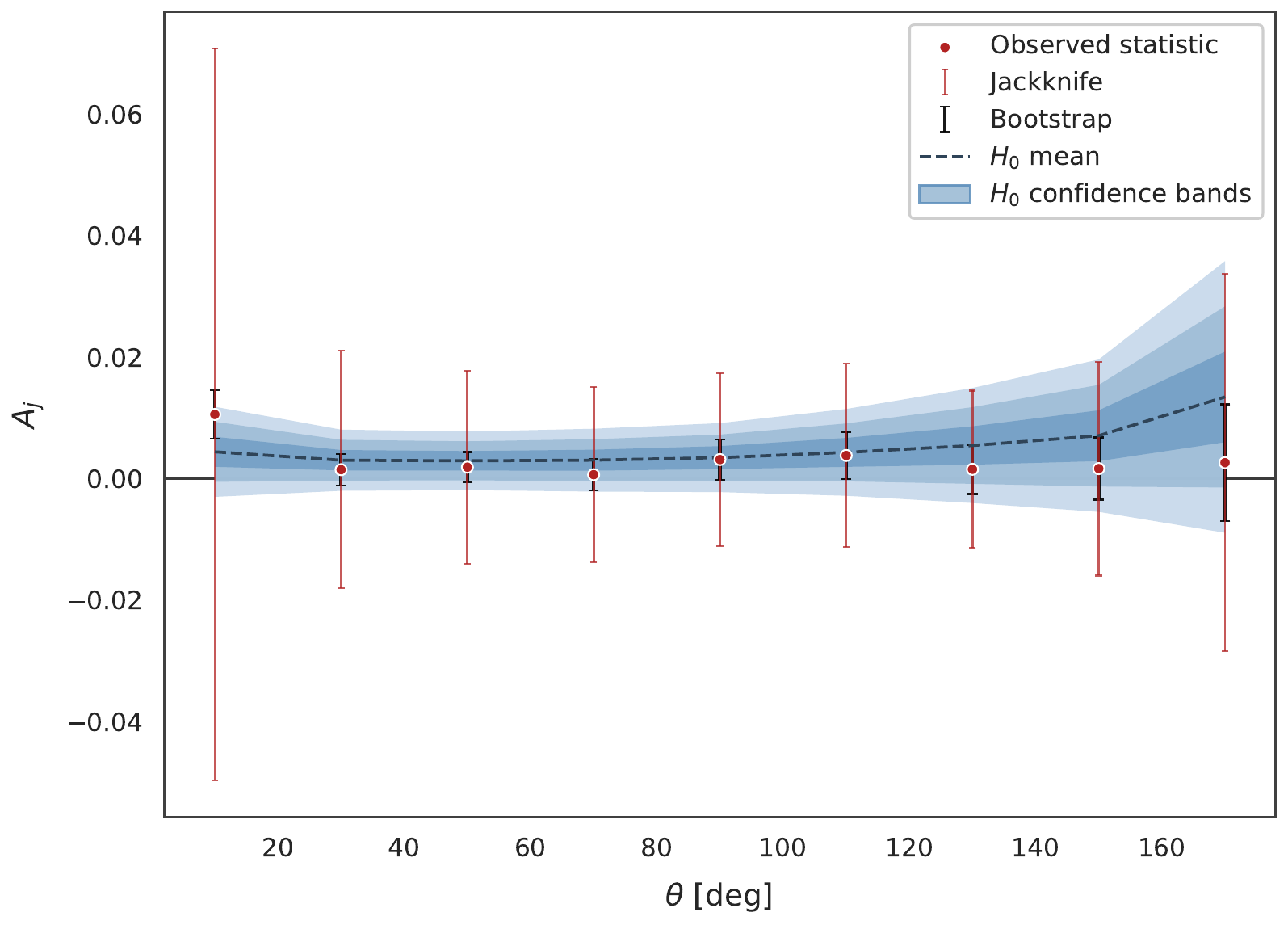}
\caption{Same as Fig.~\ref{fig:pure_abs}, but for the \emph{Fiducial} scenario. The observed $A_j$ is likewise statistically consistent with the isotropic mock ensemble.}
\label{fig:fiducial_abs}
\end{figure}

Our primary isotropy diagnostic is the SVD-regularized, empirically-calibrated significance $\sigma_{\chi^2_{\rm SVD}}$, described in Sec.~\ref{subsec:inference}. Table~\ref{tab:main-svd} reports this quantity, alongside the SVD-regularized statistic $\chi^2_{\rm SVD}$, its reduced form $\chi^2_{\rm SVD,red}$, and the analytic and empirical $p$-values, for the four isotropy scenarios. As an independent cross-check, Table~\ref{tab:abs-summary} presents the corresponding absolute-anisotropy results. The associated angular-correlation profiles are shown in Figs.~\ref{fig:pure_2pacf}--\ref{fig:fiducial_abs}. The empirical selection functions underlying the fiducial scenario are presented separately in Appendix~\ref{app:sfvalidation}. Complementary results, including the full-covariance statistics, covariance-quality diagnostics, and heuristic KS/AD tests, are collected in Appendix~\ref{app:supplementary}, where the full-covariance and SVD-regularized statistics are shown to agree to better than $1\%$ throughout.

Table~\ref{tab:main-svd} displays a clear and physically meaningful progression across the four scenarios. Masking alone reduces the SVD-regularized reduced statistic by only a factor of $\sim3$ (from $\chi^2_{\rm SVD,red}\simeq2.2\times10^5$ to $\simeq7.0\times10^4$), whereas selection-function weighting alone reduces it by a factor of $\sim1.8\times10^4$ (down to $\chi^2_{\rm SVD,red}\simeq11.7$) --- nearly four orders of magnitude more effective. This indicates that the dominant source of the raw catalog's apparent anisotropy is the highly non-uniform sky coverage of the contributing surveys, rather than Galactic-plane incompleteness. Combining both effects in the Fiducial configuration yields a further $\sim19\times$ improvement, bringing $\chi^2_{\rm SVD,red}$ to $0.62$ ($\sigma_{\chi^2_{\rm SVD}}=-0.807$), statistically consistent with a typical isotropic mock realization. We therefore find no evidence for residual large-scale anisotropy in the FRB sky distribution once both Galactic obscuration and empirical survey selection are accounted for. Notably, Galactic masking is far more effective once the selection function is applied ($\sim19\times$) than on its own ($\sim3\times$), consistent with survey coverage being the leading effect.

\begin{table}
\centering
\footnotesize
\setlength{\tabcolsep}{4.5pt}
\renewcommand{\arraystretch}{1.4}
\begin{tabular}{lcc ccccc}
\toprule
\bf Scenario & \bf Mask & \bf SF
 & $\boldsymbol{\chi^2_{\rm SVD}}$
 & $\boldsymbol{\chi^2_{\rm SVD,red}}$
 & $\boldsymbol{p^{\rm SVD}_{\rm an}}$
 & $\boldsymbol{p^{\rm SVD}_{\rm emp}}$
 & $\boldsymbol{\sigma_{\chi^2_{\rm SVD}}}$ \\
\midrule
Pure isotropy   & No  & No  & $3.66\times10^6$ & $2.15\times10^5$ & $0^{\ddagger}$ & $0.001^{\dagger}$ & $3.09^{\dagger}$ \\
Masked isotropy & Yes & No  & $1.20\times10^6$ & $7.04\times10^4$ & $0^{\ddagger}$ & $0.001^{\dagger}$ & $3.09^{\dagger}$ \\
SF-weighted     & No  & Yes & $199.14$         & $11.71$          & $4.26 \times10^{-33}$ & $0.001^{\dagger}$ & $3.09^{\dagger}$ \\
Fiducial        & Yes & Yes & $10.49$          & $0.62$           & $0.882$        & $0.790$           & $-0.807$ \\
\bottomrule
\end{tabular}
\caption{Primary isotropy diagnostic for the four scenarios: the SVD-regularized statistic $\chi^2_{\rm SVD}$ and $\chi^2_{\rm SVD,red}$, the analytic and empirical MC $p$-values, and the corresponding Gaussian-equivalent significance $\sigma_{\chi^2_{\rm SVD}}$. $^{\dagger}$Empirical $p$ at the MC resolution floor $1/(N_{\rm mocks}+1)\approx0.001$. $^{\ddagger}$Analytic $p$ underflows double-precision arithmetic.}
\label{tab:main-svd}
\end{table}
 
The absolute-anisotropy statistic in Table~\ref{tab:abs-summary} corroborates the Fiducial conclusion through an independent estimator, and exhibits the same physical progression as the covariance-aware analysis. The Gaussian tension falls only by a factor of $\sim1.7$ under Galactic masking alone (from $2433$ to $1439$), but by a factor of $\sim265$ under selection-function weighting alone (down to $9.18$), confirming that survey coverage is primarily responsible for the raw catalog's apparent anisotropy.

\begin{table}
\centering
\footnotesize
\setlength{\tabcolsep}{4.5pt}
\renewcommand{\arraystretch}{1.4}
\begin{tabular}{lccc}
\toprule
\bf Scenario & $\boldsymbol{A_{\rm obs}}$ & $\boldsymbol{p_{\rm emp}}$ & \bf tension \\
\midrule
Pure isotropy   & $14.130$  & $0.001^\dagger$ & $2433.36$ \\
Masked isotropy & $12.792$  & $0.001^\dagger$ & $1438.80$ \\
SF-weighted     & $0.110$ & $0.001^\dagger$ & $9.18$ \\
Fiducial        & $0.028$ & $0.984$         & $1.70$ \\
\bottomrule
\end{tabular}
\caption{Independent absolute-anisotropy cross-check: observed statistic $A_{\rm obs}$, its empirical MC $p$-value, and the Gaussian tension relative to the isotropic mock ensemble. Floor marker as in Table~\ref{tab:main-svd}.}
\label{tab:abs-summary}
\end{table}

Combining both effects brings the Fiducial tension to $1.70$, with $A_{\rm obs}$ well within the isotropic mock distribution ($p_{\rm emp}=0.984$). The fact that two independently constructed estimators agree so closely reinforces the robustness of the isotropy conclusion. This $p$-value indicates that the observed sky exhibits less residual angular structure than the vast majority ($\sim 98\%$) of the isotropic mock realizations, consistent with the empirically reconstructed selection function partially absorbing intrinsic angular structure.


The angular profiles in Figs.~\ref{fig:pure_2pacf}--\ref{fig:fiducial_abs} display this same progression visually. In the uncorrected scenarios (Figs.~\ref{fig:pure_2pacf}--\ref{fig:masked_abs}), $w(\theta)$ exhibits a pronounced large-scale gradient, declining monotonically from strongly positive values at small separations ($w\simeq2.5$--$2.7$) through zero near $\theta\simeq80^\circ$ to negative values ($w\simeq-0.7$) at large separations, and lies everywhere far above the essentially null isotropic benchmark denoted by the $H_0$ mean and confidence bands, which are visually suppressed in these plots, as well as the absolute statistic $A_j$ shows the corresponding excess. The Pure and Masked profiles are nearly indistinguishable, confirming visually that Galactic masking alone barely alters the signal. Once the survey selection functions are folded into the null hypothesis (Figs.~\ref{fig:weighted_2pacf}--\ref{fig:fiducial_abs}), the vertical scale is reduced by more than an order of magnitude: the observed $w(\theta)$ and $A_j$ points scatter tightly around the isotropic mean and remain within both the $H_0$ confidence bands and the resampling error bars at all separations, with the bands widening at the smallest and largest angles, where fewer pairs and mask boundaries reduce the statistical precision. This is the visual signature of a dipole-like footprint gradient, imprinted by the CHIME-dominated declination coverage, being absorbed into the observational model, leaving a residual consistent with statistical isotropy. In every figure, the jackknife error bars (red) are systematically larger than the bootstrap ones (black) --- a visual reflection of jackknife's known tendency to overestimate, and bootstrap's tendency to underestimate, the true observational variance.

\subsection{Robustness Analysis}
\label{subsec:robustness}
 
To assess whether our conclusion of consistency with statistical isotropy is sensitive to the methodological choices entering the analysis, we perform a dedicated robustness study around our fiducial configuration, whose parameter values are given in Appendix~\ref{app:fiducial}.

We vary the four methodological parameters listed in Table~\ref{tab:oat_parameters} in a one-at-a-time (OAT) sweep, changing a single parameter while the others remain at their fiducial values. Because $\sigma_{\rm smooth}$ and $N_{\rm side}^{\rm SF}$ jointly control the angular complexity of the empirical selection functions, we additionally explore their combined impact through a two-dimensional grid while keeping $b_{\rm cut}$ and $\Delta\theta$ fixed. Nominally this comprises one fiducial run, twelve OAT variations, and a $3\times3$ grid, $N_{\rm runs}=1+12+9=22$; several combinations coincide by construction, so only $13$ independent pipeline runs are required. All random seeds are held fixed across the sweep, so that differences between configurations are driven by the parameter variations rather than by independent Monte Carlo fluctuations. For each run we record the primary statistic $\sigma_{\chi^2_{\rm SVD}}$ and its displacement from the fiducial value, $\Delta\sigma_{\chi^2_{\rm SVD}}^{(r)}=\sigma_{\chi^2_{\rm SVD}}^{(r)}-\sigma_{\chi^2_{\rm SVD}}^{\rm fid}$.

\begin{table}[!ht]
\centering
\footnotesize
\setlength{\tabcolsep}{4.5pt}
\renewcommand{\arraystretch}{1.4}
\begin{tabular}{lc}
\toprule
\bf Parameter & \bf Values \\
\midrule
$b_{\rm cut}$           & $15^\circ,\,20^\circ,\,25^\circ$ \\
$\sigma_{\rm smooth}$   & $1^\circ,\,3^\circ,\,5^\circ$ \\
$N_{\rm side}^{\rm SF}$ & $16,\,32,\,64$ \\
$\Delta\theta$          & $5^\circ,\,10^\circ,\,15^\circ$ \\
\bottomrule
\end{tabular}
\caption{Methodological parameters and the values explored in the OAT robustness sweep. The fiducial value is the central entry of each row.}
\label{tab:oat_parameters}
\end{table}

\begin{table}
\centering
\footnotesize
\setlength{\tabcolsep}{4.5pt}
\renewcommand{\arraystretch}{1.4}
\begin{tabular}{cccc ccc}
\toprule
$\boldsymbol{b_{\rm cut}}$ &
$\boldsymbol{\sigma_{\rm smooth}}$ &
$\boldsymbol{N_{\rm side}^{\rm SF}}$ &
$\boldsymbol{\Delta\theta}$ &
$\boldsymbol{p_{\rm emp}}$ &
$\boldsymbol{\sigma_{\chi^2_{\rm SVD}}}$ &
$\boldsymbol{\Delta\sigma_{\chi^2_{\rm SVD}}}$
\\
\midrule
$20^\circ$ & $3^\circ$ & 32 & $10^\circ$ & 0.790 & -0.807 & 0.000\\
\midrule
$\boldsymbol{15^\circ}$ & $3^\circ$ & 32 & $10^\circ$ & 0.258 & 0.650 & +1.457\\
$\boldsymbol{25^\circ}$ & $3^\circ$ & 32 & $10^\circ$ & 0.852 & -1.046 & -0.239\\
\midrule
$20^\circ$ & $\boldsymbol{1^\circ}$ & 32 & $10^\circ$ & 0.533 & -0.084 & +0.723\\
$20^\circ$ & $\boldsymbol{5^\circ}$ & 32 & $10^\circ$ & 0.101 & 1.276 & +2.083\\
\midrule
$20^\circ$ & $3^\circ$ & $\mathbf{16}$ & $10^\circ$ & 0.635 & -0.346 & +0.461\\
$20^\circ$ & $3^\circ$ & $\mathbf{64}$ & $10^\circ$ & 0.666 & -0.430 & +0.377\\
\midrule
$20^\circ$ & $3^\circ$ & 32 & $\boldsymbol{5^\circ}$ & 0.614 & -0.291 & +0.516\\
$20^\circ$ & $3^\circ$ & 32 & $\boldsymbol{15^\circ}$ & 0.854 & -1.054 & -0.247\\
\bottomrule
\end{tabular}
\caption{OAT robustness sweep, showing the full parameter configuration of every run. The first row is the fiducial configuration; in each subsequent block the parameter shown in bold is varied to two alternative values around its fiducial value, while the other three are held fixed.}
\label{tab:robustness}
\end{table} 

\begin{figure}[tbp]
\centering
\includegraphics[width=0.8\linewidth]{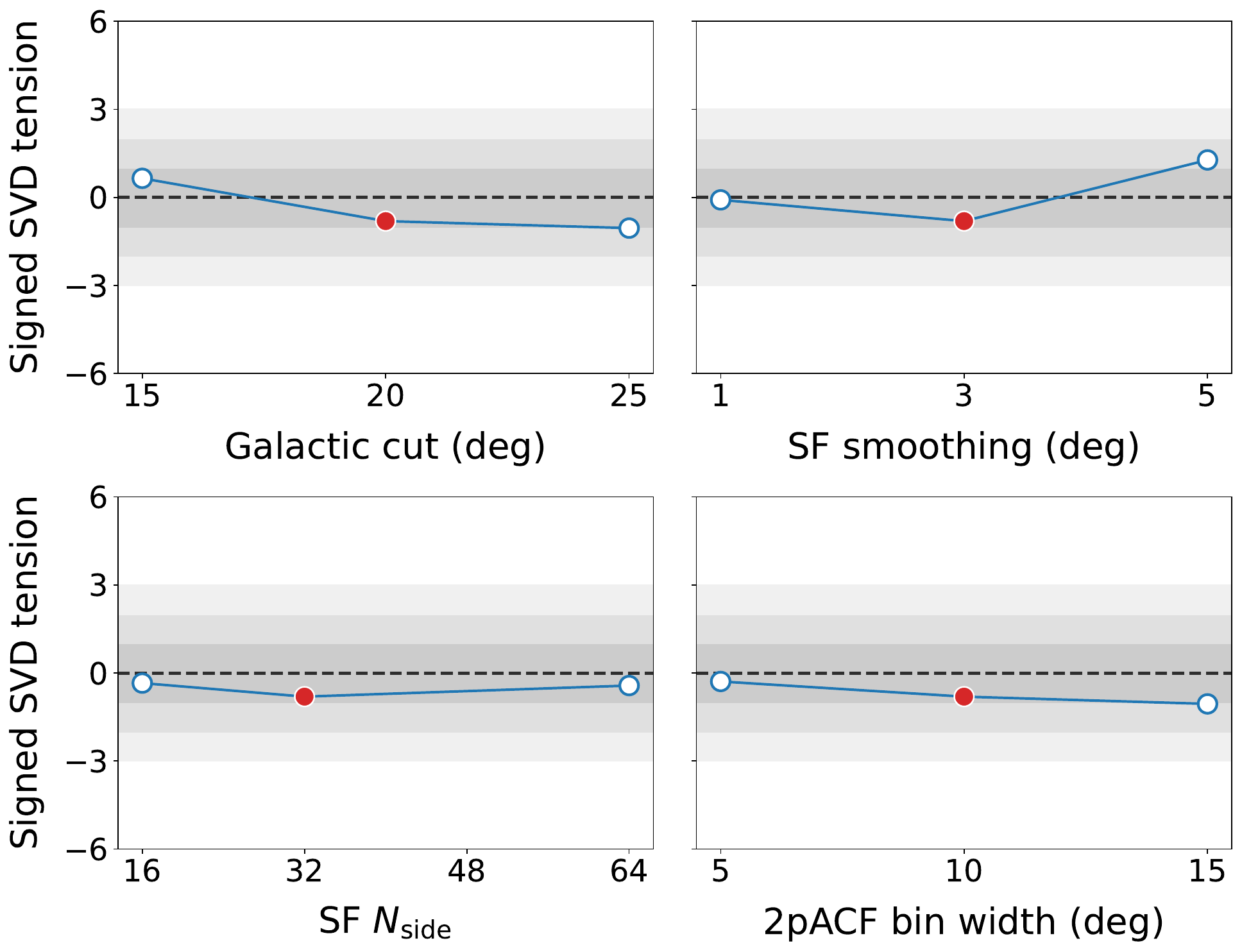}
\caption{OAT sensitivity of the primary statistic $\sigma_{\chi^2_{\rm SVD}}$ to the four methodological parameters. Each panel varies a single parameter while the remaining ones are fixed at their fiducial values. The red marker denotes the fiducial configuration.}
\label{fig:robustness}
\end{figure}
\begin{figure}[tbp]
\centering
\includegraphics[width=0.65\linewidth]{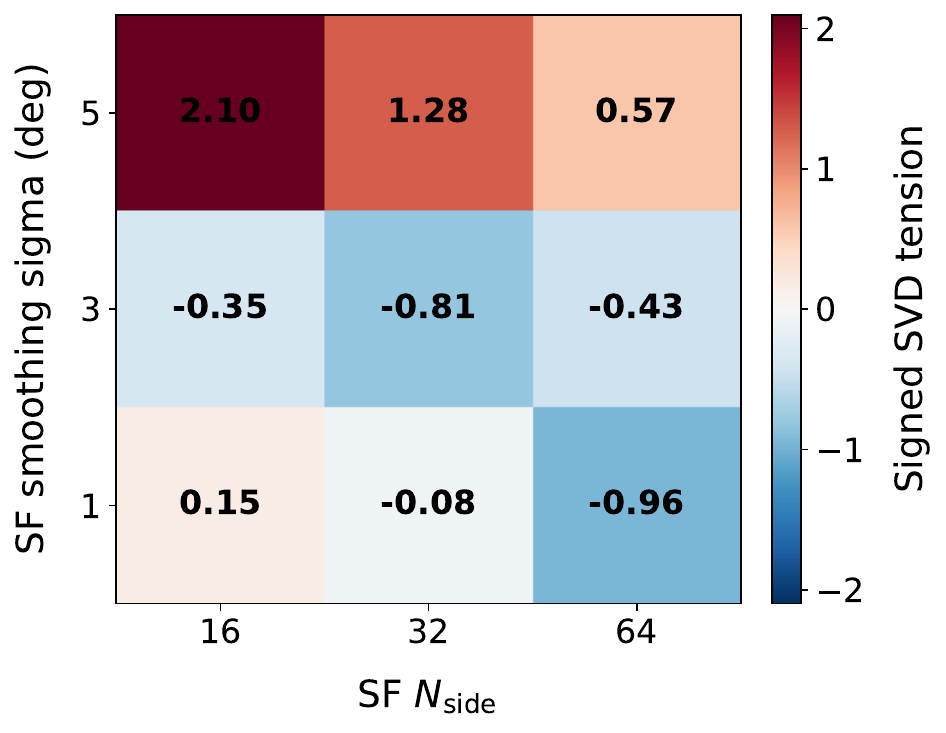}
\caption{Values of the primary statistic $\sigma_{\chi^2_{\rm SVD}}$ over the $\sigma_{\rm smooth}\times N_{\rm side}^{\rm SF}$ grid, with $b_{\rm cut}$ and $\Delta\theta$ fixed at their fiducial values.}
\label{fig:robustness2}
\end{figure}
 
Table~\ref{tab:robustness} summarizes the OAT sweep, while Figs.~\ref{fig:robustness} and~\ref{fig:robustness2} display the corresponding OAT sensitivity curves and the $\sigma_{\rm smooth}\times N_{\rm side}^{\rm SF}$ heatmap. Across all $13$ independent configurations, the qualitative conclusion is unchanged: none rejects statistical isotropy, with $\sigma_{\chi^2_{\rm SVD}}<3$ and empirical $p$-values exceeding $0.01$ throughout. The fiducial configuration yields $\sigma_{\chi^2_{\rm SVD}}=-0.807$, while all OAT configurations remain within the interval $-1.05\le\sigma_{\chi^2_{\rm SVD}}\le1.28$. Relative to the fiducial configuration, the largest OAT displacement is $|\Delta\sigma_{\chi^2_{\rm SVD}}|=2.08$, obtained for the strongest selection-function smoothing ($\sigma_{\rm smooth}=5^\circ$). Considering the full $\sigma_{\rm smooth}\times N_{\rm side}^{\rm SF}$ grid, the explored range expands to $-1.05\le\sigma_{\chi^2_{\rm SVD}}\le2.10$, with the largest excursion $\Delta\sigma_{\chi^2_{\rm SVD}}\simeq2.91$ occurring at $(\sigma_{\rm smooth},N_{\rm side}^{\rm SF})=(5^\circ,16)$ ($\sigma_{\chi^2_{\rm SVD}}=2.10$, $p_{\rm emp}=0.018$). This configuration corresponds to the smoothest and lowest-resolution selection function, for which genuine angular structure is partially washed out, mildly increasing the apparent tension, but still without approaching the $3\sigma$ rejection threshold. We therefore conclude that the qualitative isotropy result is robust, as the precise numerical significance only exhibits a moderate dependence on the adopted selection-function reconstruction only in this extreme corner of the parameter space.


\section{Conclusion}
\label{sec:conclusion}

In this work, we tested the assumption of statistical large-scale isotropy in the Universe through the angular distribution of FRBs on the celestial sphere. We used two complementary estimators: the 2pACF $w(\theta)$, measured with the Landy--Szalay estimator, and the tomographic absolute-anisotropy statistic $A_j$ (Eq.~\ref{eq:abs-sum}), which accumulates angular power, and it is more robust against cancellations between positive and negative fluctuations. Both estimators were confronted with hierarchical ensembles of isotropic mock catalogs that simultaneously propagate the uncertainties of the empirically-reconstructed selection functions and the Poisson fluctuations of the isotropic realizations. The covariance was estimated from these ensembles and stabilized through Ledoit--Wolf shrinkage, Hartlap debiasing, and an SVD truncation of its near-singular eigenmodes, and the resulting significances were calibrated empirically against the mock ensemble rather than assumed from asymptotic distributions.

Our analysis is based on the current FRB sample, a compilation of $4066$ events detected by multiple surveys. Most of them are located in the northern hemisphere, due to the CHIME contribution. Because publicly available survey information is insufficient to construct a consistent angular exposure model across all instruments, the selection functions were reconstructed directly from the observed sky distributions, i.e., following an empirical approach. We performed the analysis under four nested scenarios: \emph{Pure isotropy}, using the complete observed catalog; \emph{Masked isotropy}, after removing the Galactic plane; \emph{SF-weighted isotropy}, applying the empirical survey selection functions; and \emph{Fiducial}, combining the Galactic mask with the selection functions. All of these configurations were applied on MC realizations of objects evenly distributed across the sky, which comprises our null hypothesis here.

The four scenarios trace a clear and physically meaningful progression. Pure isotropy is largely inconsistent with the isotropic benchmark (Figs.~\ref{fig:pure_2pacf} and~\ref{fig:pure_abs}), with $\chi^2_{\rm SVD,red}\approx2\times10^5$. Removing the Galactic plane reduces this by only a factor of $\sim3$ ($\chi^2_{\rm SVD,red}\approx7\times10^4$), leaving the distribution still strongly anisotropic (Figs.~\ref{fig:masked_2pacf} and~\ref{fig:masked_abs}). Applying the empirical selection functions instead reduces it by nearly four orders of magnitude ($\chi^2_{\rm SVD,red}\approx11.7$), although this scenario on its own remains formally inconsistent with isotropy (Figs.~\ref{fig:weighted_2pacf} and~\ref{fig:weighted_abs}). Only when both effects are combined, in the Fiducial configuration, the observed distribution becomes fully consistent with the isotropic benchmark at all angular separations, as shown in Figs.~\ref{fig:fiducial_2pacf} and~\ref{fig:fiducial_abs}), with $\chi^2_{\rm SVD,red}=0.62$ and $\sigma_{\chi^2_{\rm SVD}}=-0.807$. This conclusion is independently corroborated by the absolute-anisotropy estimator ($p_{\rm emp}=0.984$). The dominant source of the raw catalog's apparent anisotropy is therefore the highly non-uniform sky coverage of the contributing surveys, overwhelmingly dominated by CHIME, rather than Galactic plane contamination and/or incompleteness.

We also explored the fiducial configuration under variations of four methodological parameters (Table~\ref{tab:oat_parameters}) to assess whether this conclusion is sensitive to such choices. None of the configurations rejects statistical isotropy (Table~\ref{tab:robustness}), showing that the qualitative result is stable against our modeling choices, even though the precise numerical significance varies under the most aggressive selection-function configurations.

As a cautious note, two caveats frame the interpretation of this result. First, because the survey selection functions are reconstructed empirically from the observed distributions themselves, they may partially absorb a genuine large-scale anisotropy, so that our test is necessarily less sensitive to signals aligned with the survey footprint; quantifying this loss of sensitivity requires a dedicated injection-recovery analysis, which we address separately. Second, the sample remains dominated by a single survey, so that the effective sky coverage, rather than the total number of detected events, currently limits the constraining power of the test. Nevertheless, we were able to recover the expected underlying distribution of FRBs, considering that they are located in host galaxies that trace the matter density distribution in the Universe, and hence must follow large-scale statistical isotropy if the Cosmological Principle is indeed a physically valid assumption to describe it.

We expect that the next generation of FRB surveys and facilities, such as the CHIME/FRB Outriggers \cite{CHIME_outriggers}, the \textit{Square Kilometre Array} (SKA-Mid and SKA-Low) \cite{SKA}, and the DSA's next-generation antennas, the DSA-2000 \cite{DSA2000}, which is designed to cover almost the entire observable sky, will significantly increase the number of detected events, especially in the southern hemisphere, where only a few detections are currently available. With a larger and more uniformly distributed sample, we expect a reduction in statistical uncertainties and, more importantly, a weakening of the degeneracy between a genuine cosmological signal and the observational response that limits the present analysis. As the number of FRB detections continues to grow, these events will provide increasingly robust tests of cosmic isotropy and further demonstrate the potential of FRBs as high-precision cosmological probes.


\acknowledgments

BWNR, ARQ, KELF and LLS thank the Fundação de Apoio à Pesquisa do Estado da Paraíba (FAPESQ). TL and KELF thank the financial support from the Conselho Nacional de Desenvolvimento Científico e Tecnológico (CNPq). CAPB acknowledges financial support from the CNPq grant 306630/2025-7. ARQ acknowledges the financial support by CNPq under process number 306884/2026-7. JSA is supported by Conselho Nacional de Desenvolvimento Cient\'{\i}fico e Tecnol\'ogico (CNPq No. 307683/2022-2; CNPq No. 448158/2025-6) and Funda\c{c}\~ao de Amparo \`a Pesquisa do Estado do Rio de Janeiro (FAPERJ) grant 259610 (2021). K.E.L.F. also thanks FAPESQ for the exchange program Paraíba sem Fronteiras and the ENSEMBLE3 project, which is carried out within the 2.1 International Research Agendas programme of the Foundation for Polish Science co-financed by the European Union under the European Funds for Smart Economy 2021-2027 (FENG.02.01-IP.05-0044/24), project (MAB/2020/14), which is carried out within the International Research Agendas Programme (IRAP) of the Foundation for Polish Science co-financed by the European Union under the European Regional Development Fund and the Teaming Horizon 2020 programme (GA. No. 857543) of the European Commission and the project of the Minister of Science and Higher Education "Support for the activities of Centers of Excellence established in Poland under the Horizon 2020 program" under contract No. MEiN/2023/DIR/3797.


\appendix

\section{Fiducial Analysis Parameters}
\label{app:fiducial}

The fiducial configuration adopted throughout this work is the isotropic null hypothesis $H_0$ propagated through both the Galactic mask and the empirical survey selection functions. Its complete specification is given in Table~\ref{tab:fiducial-params}.

\begin{table}[!ht]
\centering
\footnotesize
\setlength{\tabcolsep}{4.5pt}
\renewcommand{\arraystretch}{1.4}
\begin{tabular}{lcl}
\toprule
\bf Parameter & \bf Symbol & \bf Fiducial value \\
\midrule
Galactic mask & --- & Yes \\
Selection functions & --- & Yes \\
\midrule
Galactic latitude cut & $b_{\rm cut}$ & $20^\circ$ \\
SF HEALPix resolution & $N_{\rm side}^{\rm SF}$ & $32$ \\
SF smoothing scale & $\sigma_{\rm smooth}$ & $3^\circ$ \\
JK HEALPix resolution & $N_{\rm side}^{\rm JK}$ & $4$ \\
2pACF bin width & $\Delta\theta$ & $10^\circ$ \\
Number of fine 2pACF bins & $N_{\rm bins}$ & $18$ \\
Number of coarse tomographic bins & $N_{\rm tomo}$ & $9$ \\
SF perturbation realizations & $N_{\rm ens}$ & $20$ \\
Mocks per SF realization & $N_{\rm mocks}^{\rm ens}$ & $50$ \\
Total isotropic mocks & $N_{\rm mocks}$ & $1000$ \\
Random-to-data ratio & $N_{\rm rand}/N_{\rm data}$ & $20$ \\
Bootstrap realizations & $N_{\rm BS}$ & $500$ \\
SVD eigenvalue cut & --- & $10^{-2}\lambda_{\max}$ \\
\bottomrule
\end{tabular}
\caption{Complete specification of the fiducial configuration adopted in the primary analysis: the two observational ingredients (top) and the methodological parameter values (bottom).}
\label{tab:fiducial-params}
\end{table}

\section{SF Validations and Survey Maps}
\label{app:sfvalidation}

\begin{figure}
    \centering
    \includegraphics[width=0.9\linewidth]{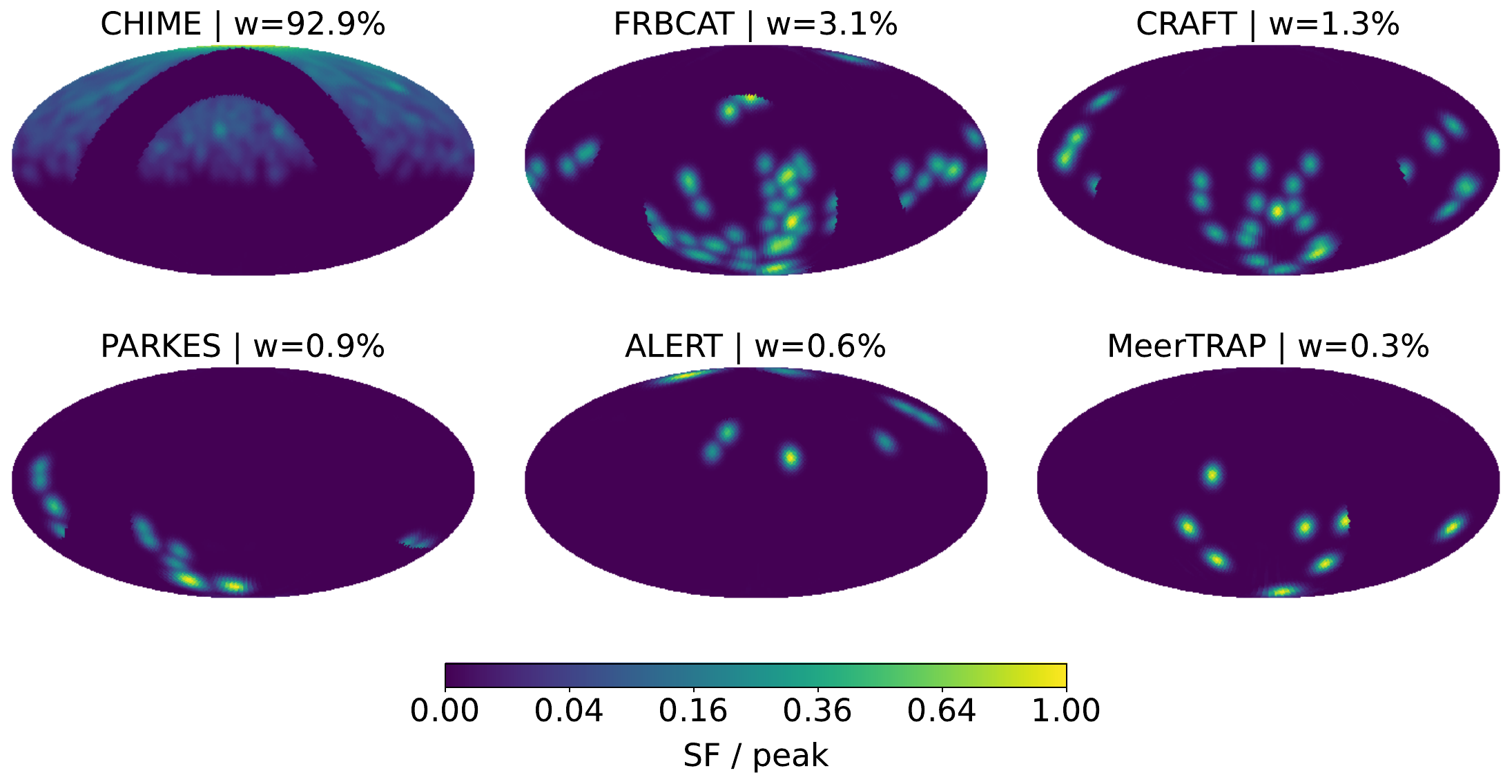}
    \caption{Reconstructed selection-function maps ${\rm SF}(\hat n)$ for the six highest-weight surveys, in equatorial (ICRS) coordinates.}
\label{fig:valid}
\end{figure}

\begin{figure}
\centering
\includegraphics[width=0.9\linewidth]{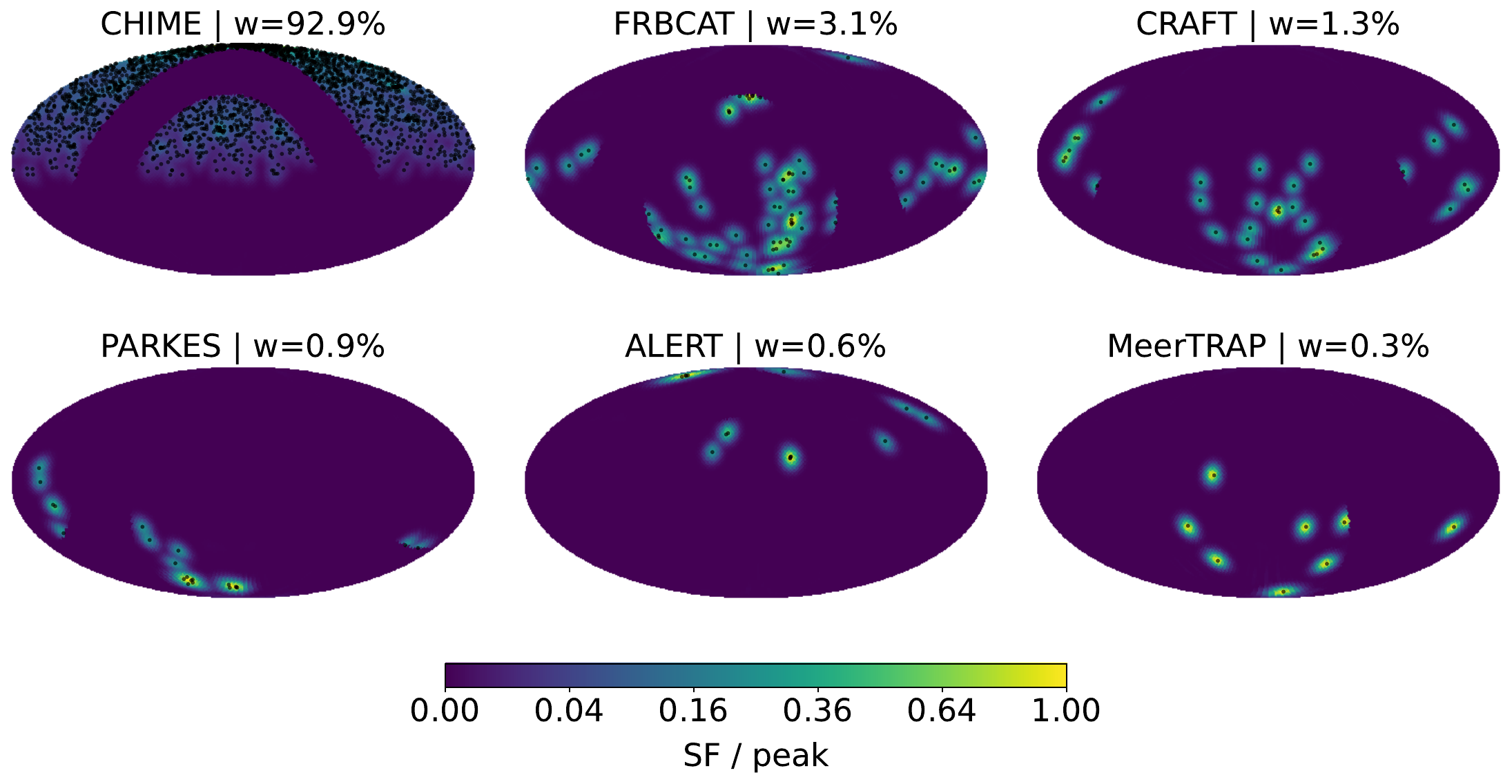}
\caption{Same selection-function maps as in Fig.~\ref{fig:valid}, with the observed FRB positions (black dots) of each survey overlaid.}
\label{fig:svmaps}
\end{figure}

The reconstructed selection functions ${\rm SF}(\hat n)$, after smoothing, masking, and renormalization, underlying the fiducial analysis are shown for the six highest-weighted surveys, ordered by decreasing weight $w_i$. Fig.~\ref{fig:valid} shows the maps individually; Fig.~\ref{fig:svmaps} overlays the observed FRB positions of each survey, confirming that the reconstruction traces the actual data footprint. Each panel is normalized to its own peak value (${\rm SF}/{\rm peak}$) for display only, which does not affect the selection functions used in the analysis; a square-root stretch is applied to the shared color scale to enhance the visibility of broad, low-amplitude coverage (e.g., CHIME's near-uniform footprint) relative to the sharply peaked coverage of smaller surveys.

\section{Supplementary Statistics and Covariance Diagnostics}
\label{app:supplementary}

This appendix collects the statistics that support the SVD-regularized empirical diagnostic reported in Sec.~\ref{sec:results}: the full-covariance $\chi^2$ statistics and analytic $p$-values, the covariance-quality diagnostics, and the heuristic KS/AD tests.

Table~\ref{tab:full-cov} lists the full-covariance statistics and their analytic and empirical significances for the four scenarios. The two estimators agree to better than $1\%$ in $\chi^2$ throughout (compare Tables~\ref{tab:main-svd} and~\ref{tab:full-cov}), so that the SVD regularization stabilizes the covariance inversion without distorting the inferred significance. The larger offset between the reduced values reflects only the different normalization, $N_{\rm bins}=18$ against $N_{\rm kept}=17$.
\begin{table}[!ht]
\centering
\footnotesize
\setlength{\tabcolsep}{4.5pt}
\renewcommand{\arraystretch}{1.4}
\begin{tabular}{lcc ccccc}
\toprule
\bf Scenario & \bf Mask & \bf SF
 & $\boldsymbol{\chi^2}$
 & $\boldsymbol{\chi^2_{\rm red}}$
 & $\boldsymbol{p_{\rm an}}$ & $\boldsymbol{p_{\rm emp}}$ & $\boldsymbol{\sigma_{\chi^2}}$ \\
\midrule
Pure isotropy   & No  & No  & $3.66\times10^6$ & $2.03\times10^5$ & $0^{\ddagger}$ & $0.001^{\dagger}$ & $3.09^{\dagger}$ \\
Masked isotropy & Yes & No  & $1.20\times10^6$ & $6.64\times10^4$ & $0^{\ddagger}$ & $0.001^{\dagger}$ & $3.09^{\dagger}$ \\
SF-weighted     & No  & Yes & $198.96$         & $11.05$          & $1.62\times10^{-32}$ & $0.001^{\dagger}$ & $3.09^{\dagger}$ \\
Fiducial        & Yes & Yes & $10.58$          & $0.59$           & $0.911$        & $0.789$           & $-0.804$ \\
\bottomrule
\end{tabular}
\caption{Full-covariance $\chi^2$ statistics and significances for the four scenarios, complementary to the SVD-regularized values of Table~\ref{tab:main-svd}. $^{\dagger}$$p$-value at the MC resolution floor. $^{\ddagger}$Analytic $p$ underflows double-precision arithmetic.}
\label{tab:full-cov}
\end{table}

This agreement holds at the level of the calibrated $p$-values as well. For the Fiducial scenario the four available calibrations return $p_{\rm an}=0.911$ and $p_{\rm emp}=0.789$ under the full covariance, against $p^{\rm SVD}_{\rm an}=0.882$ and $p^{\rm SVD}_{\rm emp}=0.790$ under the regularized one: the two empirical values differ by $0.001$, the resolution of the mock ensemble $1/(N_{\rm mocks}+1)$, and the two analytic ones by only $0.029$. The analytic values sit slightly above their empirical counterparts, as expected, since they assume an exact $\chi^2$ distribution whereas the empirical calibration inherits the residual correlations and non-Gaussianity encoded in the mock ensemble; all four nonetheless agree that the Fiducial catalog is comfortably consistent with statistical isotropy.

Table~\ref{tab:cov-diagnostics} reports the quality diagnostics of the covariance matrices. The Hartlap correction factors are close to unity ($\alpha\simeq0.98$) and $N_{\rm kept}=17$ of $18$ eigenmodes survive the SVD cut in every scenario. The condition number is reduced by a factor of $\sim20$ in all four cases ($\kappa_{\rm SVD}\ll\kappa_{\rm full}$): the SVD truncation removes the near-singular directions responsible for the instability of the full-covariance inverse, which is what makes the regularized statistic numerically reliable.

\begin{table}
\centering
\footnotesize
\setlength{\tabcolsep}{4.5pt}
\renewcommand{\arraystretch}{1.4}
\begin{tabular}{lcccccc}
\toprule
 & \multicolumn{2}{c}{\bf Hartlap corrections}
 & \multicolumn{2}{c}{\bf SVD truncation}
 & \multicolumn{2}{c}{\bf Cov. cond. numb.} \\
\cmidrule(lr){2-3}
\cmidrule(lr){4-5}
\cmidrule(lr){6-7}
\bf Scenario
& $\boldsymbol{\alpha_{\rm full}}$
& $\boldsymbol{\alpha_{\rm SVD}}$
& $\boldsymbol{N_{\rm kept}/N_{\rm bins}}$
& $\boldsymbol{N_{\rm eff}}$
& $\boldsymbol{\kappa_{\rm full}}$
& $\boldsymbol{\kappa_{\rm SVD}}$ \\
\midrule
Pure isotropy   & $0.981$ & $0.982$ & $17/18$ & $7.6$ & $288.7$  & $12.6$ \\
Masked isotropy & $0.981$ & $0.982$ & $17/18$ & $9.5$ & $169.5$  & $8.0$ \\
SF-weighted     & $0.981$ & $0.982$ & $17/18$ & $3.7$ & $1363.7$ & $58.0$ \\
Fiducial        & $0.981$ & $0.982$ & $17/18$ & $4.6$ & $956.7$  & $45.1$ \\
\bottomrule
\end{tabular}
\caption{Quality diagnostics of the covariance matrices: Hartlap correction factors, retained eigenmodes after SVD truncation, effective number of mocks, and covariance condition numbers before and after SVD regularization.}
\label{tab:cov-diagnostics}
\end{table}

\begin{table*}
\centering
\footnotesize
\setlength{\tabcolsep}{4.5pt}
\renewcommand{\arraystretch}{1.4}
\begin{tabular}{lcccccccc}
\toprule
\multirow{2}{*}{\bf Scenario}
& \multicolumn{2}{c}{\bf KS ($\boldsymbol{w}$)}
& \multicolumn{2}{c}{\bf AD ($\boldsymbol{w}$)}
& \multicolumn{2}{c}{\bf KS ($\boldsymbol{A_j}$)}
& \multicolumn{2}{c}{\bf AD ($\boldsymbol{A_j}$)} \\
\cmidrule(lr){2-3}
\cmidrule(lr){4-5}
\cmidrule(lr){6-7}
\cmidrule(lr){8-9}
& $\boldsymbol{p_{\rm an}}$
& $\boldsymbol{p_{\rm emp}}$
& $\boldsymbol{p_{\rm an}}$
& $\boldsymbol{p_{\rm emp}}$
& $\boldsymbol{p_{\rm an}}$
& $\boldsymbol{p_{\rm emp}}$
& $\boldsymbol{p_{\rm an}}$
& $\boldsymbol{p_{\rm emp}}$ \\
\midrule
Pure isotropy
& $0.007$ & $0.219$
& $0.003$ & $0.095$
& $4.11\times10^{-5}$ & $0.002$
& $0.001^{\S}$ & $0.002$ \\

Masked isotropy
& $0.007$ & $0.333$
& $0.003$ & $0.210$
& $4.11\times10^{-5}$ & $0.001^{\dagger}$
& $0.001^{\S}$ & $0.001^{\dagger}$ \\

SF-weighted
& $0.021$ & $0.794$
& $0.004$ & $0.262$
& $0.126$ & $0.107$
& $0.020$ & $0.007$ \\

Fiducial
& $0.0004$ & $0.013$
& $0.002$ & $0.012$
& $0.034$ & $0.040$
& $0.020$ & $0.011$ \\
\bottomrule
\end{tabular}
\caption{Analytic and empirical KS and AD diagnostics for the $w(\theta)$ and $A_j$ profiles in the four scenarios. Floor markers as in Table~\ref{tab:full-cov}; $^{\S}$analytic AD $p$ at \textsc{SciPy}'s tabulated floor.}
\label{tab:nonparametric-results}
\end{table*}

Table~\ref{tab:nonparametric-results} reports the heuristic KS and AD tests for the $w(\theta)$ and $A_j$ profiles, and illustrates why they are used only as qualitative cross-checks. For the three scenarios that the covariance-aware statistic rejects outright ($\sigma_{\chi^2_{\rm SVD}}=3.09$), the pointwise KS test applied to $w(\theta)$ nonetheless reports no tension whatsoever ($p_{\rm emp}=0.219$, $0.333$, and $0.794$ for Pure isotropy, Masked isotropy, and SF-weighted, respectively): KS is sensitive only to the single most discrepant point of the empirical distribution and cannot coherently accumulate evidence spread across correlated angular bins, whereas $\chi^2_{\rm SVD}$ is explicitly constructed to do so. The same scenarios are, by contrast, correctly flagged by both tests when applied to the tomographic $A_j$ profile, where the departure is concentrated in fewer bins. The converse failure appears for the Fiducial scenario, where all four test/profile combinations return low empirical $p$-values ($p_{\rm emp}\simeq0.011$--$0.040$) despite the fully consistent covariance-aware result; given the heuristic nature of these tests under correlated bins, we interpret this as a correlation-driven artifact rather than genuine residual anisotropy. We therefore adopt $\sigma_{\chi^2_{\rm SVD}}$ as the statistically preferred diagnostic throughout.


\bibliographystyle{JHEP}
\bibliography{jcapbib}

\end{document}